\documentclass[12pt]{article}
\usepackage[inline,shortlabels]{enumitem}
\usepackage{float}
\usepackage{ulem}
\usepackage{mathtools}
\usepackage{graphicx}
\usepackage[round]{natbib}
\usepackage[margin=0.87in]{geometry}
\usepackage{pgf,tikz}
\usepackage[english]{babel}
\usepackage{longtable}
\usepackage{xcolor}
\usepackage{amssymb,amsmath,amsthm}
\usepackage{multirow}
\usepackage[titletoc,title]{appendix}
\usepackage{authblk}
\usepackage{setspace}
\usepackage{dsfont}
\usepackage[OT1]{fontenc}
\usepackage{subcaption}
\usepackage{refcount}
\usepackage{thmtools}
\usepackage{slashed}
\usetikzlibrary{arrows.meta}
\usetikzlibrary{decorations.markings}
\usepackage[linesnumbered]{algorithm2e}
\RestyleAlgo{ruled}

\graphicspath{ {./figs/} }
\usetikzlibrary{arrows,shapes.arrows,shapes.geometric,shapes.multipart,
  decorations.pathmorphing,positioning}
\usepackage{xr-hyper}
\usepackage[colorlinks,citecolor=magenta,urlcolor=magenta]{hyperref}
\declaretheorem[style=definition]{example}
\makeatletter
\let\save@mathaccent\mathaccent
\newcommand*\if@single[3]{%
  \setbox0\hbox{${\mathaccent"0362{#1}}^H$}%
  \setbox2\hbox{${\mathaccent"0362{\kern0pt#1}}^H$}%
  \ifdim\ht0=\ht2 #3\else #2\fi
}
\renewcommand{\emph}[1]{\textit{#1}}
\newcommand*\rel@kern[1]{\kern#1\dimexpr\macc@kerna}
\newcommand*\widebar[1]{\@ifnextchar^{{\wide@bar{#1}{0}}}{\wide@bar{#1}{1}}}
\newcommand\hlrev[1]{\textcolor{black}{#1}}
\newcommand*\wide@bar[2]{\if@single{#1}{\wide@bar@{#1}{#2}{1}}{\wide@bar@{#1}{#2}{2}}}
\newcommand*\wide@bar@[3]{%
  \begingroup
  \def\mathaccent##1##2{%
    \let\mathaccent\save@mathaccent
    \if#32 \let\macc@nucleus\first@char \fi
    \setbox\z@\hbox{$\macc@style{\macc@nucleus}_{}$}%
    \setbox\tw@\hbox{$\macc@style{\macc@nucleus}{}_{}$}%
    \dimen@\wd\tw@
    \advance\dimen@-\wd\z@
    \divide\dimen@ 3
    \@tempdima\wd\tw@
    \advance\@tempdima-\scriptspace
    \divide\@tempdima 10
    \advance\dimen@-\@tempdima
    \ifdim\dimen@>\z@ \dimen@0pt\fi
    \rel@kern{0.6}\kern-\dimen@
    \if#31
    \overline{\rel@kern{-0.6}\kern\dimen@\macc@nucleus\rel@kern{0.4}\kern\dimen@}%
    \advance\dimen@0.4\dimexpr\macc@kerna
    \let\final@kern#2%
    \ifdim\dimen@<\z@ \let\final@kern1\fi
    \if\final@kern1 \kern-\dimen@\fi
    \else
    \overline{\rel@kern{-0.6}\kern\dimen@#1}%
    \fi
  }%
  \macc@depth\@ne
  \let\math@bgroup\@empty \let\math@egroup\macc@set@skewchar
  \mathsurround\z@ \frozen@everymath{\mathgroup\macc@group\relax}%
  \macc@set@skewchar\relax
  \let\mathaccentV\macc@nested@a
  \if#31
  \macc@nested@a\relax111{#1}%
  \else
  \def\gobble@till@marker##1\endmarker{}%
  \futurelet\first@char\gobble@till@marker#1\endmarker
  \ifcat\noexpand\first@char A\else
  \def\first@char{}%
  \fi
  \macc@nested@a\relax111{\first@char}%
  \fi
  \endgroup
}
\makeatother

\usepackage[inline]{trackchanges}
\addeditor{Ivan}
\addeditor{Nima}
\addeditor{Kara}

\newtheorem{remark}{Remark}

\newtheorem{property}{}
\newtheorem{proposition}{Proposition}
\newtheorem{theorem}{Theorem}
\AtEndDocument{\refstepcounter{theorem}\label{finalthm}}
\AtEndDocument{\refstepcounter{equation}\label{finaleq}}

{
  \theoremstyle{definition}
  
}
{
  \theoremstyle{definition}
  \newtheorem{assumptioniden}{}
}

\AtEndDocument{\refstepcounter{lemma}\label{finallemma}}
\newtheorem{coro}{Corollary}
\newtheorem{definition}{Definition}

\renewcommand{\P}{\mathsf{P}}
\newcommand{\F}{\mathsf{F}}

\newcommand{\A}{{\underline A}}
\newcommand{\p}{\mathsf{p}}
\newcommand{\m}{\mathsf{m}}
\newcommand{\ate}{\mathsf{ATE}}

\newcommand{\g}{\mathsf{g}}
\newcommand{\G}{\mathsf{G}}

\newcommand{\indep}{\mbox{$\perp\!\!\!\!\perp$}}

 \newcommand{\dd}{\mathrm{d}}

\newcommand{\E}{\mathsf{E}}

\newcommand{\cov}{\mathsf{Cov}}

\renewenvironment{proof}{{\it Proof }}{\qed \\}

\DeclarePairedDelimiterX{\norm}[1]{\lVert}{\rVert}{#1}

\pgfdeclarelayer{background}
\pgfsetlayers{background,main}
\usetikzlibrary{arrows,positioning}
\tikzset{
>=stealth',
punkt/.style={
rectangle,
rounded corners,
draw=black, very thick,
text width=6.5em,
minimum height=2em,
text centered},
pil/.style={
->,
thick,
shorten <=2pt,
shorten >=2pt,}
}
\newcommand{\Vertex}[2]
{\node[minimum width=0.6cm,inner sep=0.05cm] (#2) at (#1) {$\footnotesize#2$};
}
\newcommand{\Vertexr}[2]
{\node[rectangle, draw, minimum width=0.6cm,inner sep=0.05cm] (#2) at (#1) {$\footnotesize#2$};
}
\newcommand{\ArrowR}[3]%
{ \begin{pgfonlayer}{background}
\draw[->,#3] (#1) to[bend right=30] (#2);
\end{pgfonlayer}
}
\newcommand{\ArrowL}[3]%
{ \begin{pgfonlayer}{background}
\draw[->,#3] (#1) to[bend left=45] (#2);
\end{pgfonlayer}
}
\newcommand{\EdgeR}[3]%
{ \begin{pgfonlayer}{background}
    \draw[dashed,#3] (#1) to[bend right=30] (#2);
  \end{pgfonlayer}
}
\newcommand{\EdgeL}[3]%
{ \begin{pgfonlayer}{background}
    \draw[dashed,#3] (#1) to[bend left=30] (#2);
  \end{pgfonlayer}
}
\newcommand{\Arrow}[3]%
{ \begin{pgfonlayer}{background}
\draw[->,#3] (#1) -- +(#2);
\end{pgfonlayer}
}

\newcommand{\titlepaper}{Non-agency interventions for causal mediation
  in the presence of intermediate confounding} \date{\today}

\ifx\isblinded\undefined
\newcommand{\github}{https://github.com/idiazst/causal_influence}
\author[1]{Iv\'an D\'iaz \thanks{corresponding author:
    ivan.diaz@nyu.edu}}
\affil[1]{\small Division of Biostatistics, Department of Population
  Health\\ New York University Grossman School of Medicine\\New York, USA.}
\else
\if\isblinded1
\author{\vspace{-20mm}}
\newcommand{\github}{blinded for review}
\else
\newcommand{\github}{https://github.com/idiazst/causal_influence}
\author[1]{Iv\'an D\'iaz \thanks{corresponding author:
    ivan.diaz@nyu.edu}}
\affil[1]{\small Division of Biostatistics, Department of Population
  Health, New York University Grossman School of Medicine.}
\fi
\fi

\ifx\arxiv\undefined
\else
\fi

\allowdisplaybreaks
\renewcommand\thmcontinues[1]{Continued}

\AtEndDocument{\refstepcounter{theorem}\label{finalthm}}
\AtEndDocument{\refstepcounter{equation}\label{finaleq}}
\AtEndDocument{\refstepcounter{assumptioniden}\label{finalass}}

\title{\titlepaper}
\begin{document}
\maketitle

\begin{abstract}
  Recent approaches to causal inference have focused on causal effects
  defined as contrasts between the distribution of counterfactual
  outcomes under hypothetical interventions on the nodes of a
  graphical model.  In this article we develop theory for causal
  effects defined with respect to a different type of intervention,
  one which alters the information propagated through the edges of the
  graph. These information transfer interventions may be more useful
  than node interventions in settings in which causes are
  non-manipulable, for example when considering race or genetics as a
  causal agent. Furthermore, information transfer interventions allow
  us to define path-specific decompositions which are identified in
  the presence of treatment-induced mediator-outcome confounding, a
  practical problem whose general solution remains elusive. We prove
  that the proposed effects provide valid statistical tests of
  mechanisms, unlike popular methods based on randomized interventions
  on the mediator. We propose efficient non-parametric estimators for
  a covariance version of the proposed effects, using data-adaptive
  regression coupled with semi-parametric efficiency theory to address
  model misspecification bias while retaining $\sqrt{n}$-consistency
  and asymptotic normality. We illustrate the use of our methods in
  two examples using publicly available data.
\end{abstract}

\section{Introduction}

Statistical causal inference is primarily concerned with quantifying
the strength of causal relations through so-called causal effects,
which are defined as contrasts between distributions under
hypothetical modifications to the data generating mechanisms. These
hypothetical modifications are often called \textit{interventions},
and the variables resulting from the interventions are often called
\textit{counterfactuals} \citep{Woodward2003}. Counterfactual causal
effects can be classified into two types according to the
interventions that define them: \textit{agency} and
\textit{non-agency} causal effects. Agency causal effects are defined
with respect to interventions that can be carried out in the real
world, at least in principle. For example, a clinical researcher may
be interested in evaluating mortality rates in a hypothetical world
where all patients are given a certain treatment, and compare them
with mortality rates in a hypothetical world where no patient is given
the treatment, resulting in the so-called average treatment effect. On
the other hand, non-agency causal effects do not require that
interventions are feasible. For example, a clinical researcher may be
interested in evaluating the strength of the causal relation between a
biomarker and mortality through a hypothetical increase of one unit in
the biomarker, without regards to whether or how this increase can be
brought about. Analyses involving agency causal effects often have the
goal of recommending one of the actions under evaluation, and are
therefore of high importance to decision makers. This has led to some
philosophers \citep[e.g.,][]{menzies1993causation} and statisticians
\citep[e.g.,][]{holland1986statistics,hernan2008does,hernan2016does}
to hold the view that causal inference is only useful if it evaluates
agency effects,\footnote{This view is often called ``the
  interventionist view of causality''
  \citep[e.g.,][]{robins2022interventionist}, although that
  designation is a misnomer since all counterfactual views of
  causality, whether they are agency or non-agency, involve
  interventions \citep{Woodward2003}. A more appropriate label would
  be ``the agency view of causality'' \citep{menzies1993causation}.}
although that view has been vigorously contested, both in philosophy
\citep[e.g., ][]{Woodward2003} as well as in statistics and related
disciplines \citep[e.g.,
][]{glymour2014commentary,krieger2016tale,daniel2016commentary,
  vandenbroucke2016causality, glymour2017evaluating, pearl2018does,
  pearl2019interpretation, pearl2019sufficient}.

In this article we propose a type of non-agency causal effect based on
so-called \textit{information interventions}, which are heuristically
defined as modifications to the information transferred through the
edges of a causal graph. This view of causality as transfer of
information has been advocated in philosophy
\citep[e.g.,][]{collier1999causation,illari2014causality,
  hausman2004manipulation,malinsky2018intervening,
  weinberger2021signal} and has been previously used in statistics
\citep[e.g.,][]{janzing2013quantifying, schwartz2016causal,
  pearl2018does, pearl2019interpretation,heyang2019info,
  pearl2019sufficient,gong2021path,scholkopf2022causality}, although
causal effects defined using interventions on the information
transferred along the edges are far less common than effects that use
interventions on the nodes of the causal graph.\footnote{Following the
  cited literature, we use the phrase ``information transfer'' to
  refer to the interventions used in the paper. However, our use of
  the word information does not necessarily correspond its use in the
  field of ``information theory''.} In order to distinguish between
measures of causality defined through interventions on the nodes vs
the edges of a causal graph, we use the expression \textit{causal
  effect} to refer to the former and the expression \textit{causal
  influence} to refer to the latter.

The information interventions we develop have roots in
\textit{stochastic interventions}
\citep[e.g.,][]{korb2004varieties, robins2004effects,
  didelez2006direct,eberhardt2007interventions,diaz2013assessing,
  young2014identification,chaves2015information}, which are defined as
interventions where the nodes of a causal graph are intervened on and
replaced by random draws from user-given distributions. Examples of
stochastic interventions include incremental propensity score
interventions \citep{kennedy2019nonparametric, wen2021intervention},
and interventions that shift the exposure distribution in an additive
or multiplicative scale
\citep{Diaz12,diaz2020causal,diaz2021nonparametric,
  diaz2022causal}.  


Although we use ideas from the literature on stochastic interventions,
we depart from that literature in that we are not interested in
intervening on the nodes of the graph, but merely use random draws
from a distribution to aid in describing functional dependencies
between variables in a structural equation system. We propose to use
stochastic interventions as a means to define operations on a causal
model that remove or emulate the information transferred by a node
along certain edges of interest, while leaving the node
unaffected. Measures of causal influence thus defined can be
interpreted outside of an agency view of causality, as measures of
counterfactual functional dependence. In other words, the causal
influence measures we propose may be considered of a
\textit{descriptive} rather than a \textit{prescriptive} nature in the
sense that they unveil causal relations prevalent in the world, but do
not tell us about actions that would yield different outcomes.


Non-agency interventions may be useful for studying causes that are
non-manipulable or for which manipulations are difficult to
conceptualize or of no substantive interest. A common characteristic
of those causes (e.g., race, gender, genetics) is that their effect is
often mediated by other attributes which are more amenable to
manipulation by intervention. For example, the causal relation between
racial discrimination and health disparities in the US population is
mediated by socioeconomic factors
\citep{williams2000understanding}. Likewise, the causal relation
between polygenic risk scores and individual traits or disease is
often mediated by physiological or environmental processes that occur
in childhood or adulthood. \citep[see e.g.,][]{elkrief2021independent,
  yang2022longer}. Thus, the eventual development of feasible actions
to bring about changes (e.g., to reduce discrimination or to mitigate
genetic risks) requires an a-priori understanding of the intervening
mechanisms. This understanding can be achieved using mediation or path
analysis. Natural direct and indirect effects
\citep{RobinsGreenland92,pearl2001direct} are a popular approach for
mediation analysis, but they are not identified in the presence of
treatment-induced mediator-outcome confounding
\citep{avin2005identifiability}. A solution to this problem which has
gained traction in the applied literature is the use of randomized
interventional effects \citep{vanderweele2014effect}. Recent research
\citep{miles2022causal} has uncovered an important limitation of these
effects, namely that they fail to satisfy the \textit{sharp null
  mediational criterion}, meaning that the effect through the mediator
can be non-zero even when there is no structural relation that
operates through the mediator for any individual in the population. In
this paper we demonstrate how information interventions can be used to
define novel path-specific causal influence measures. We prove that
the proposed measures have an important property that they satisfy
appropriately defined path-specific sharp null criteria, meaning that
they take their null value when there is no mechanism operating
through the path under consideration. This means that the proposed
path-specific measures of causal influence can be used to identify and
estimate the strength of causal mechanisms, even in the presence of
mediator-outcome confounders affected by
treatment. 

The paper is organized as follows. 
In \S\ref{sec:notation} we introduce the notation and the causal model
used in the paper. In \S\ref{sec:total} we introduce the novel measure
of causal influence based on non-agency information transfer
interventions. In \S\ref{sec:pathanal} we discuss a path-analysis
method that uses these interventions, and we prove that this method
satisfies appropriately defined path-specific sharp null criteria,
providing correct tests for mechanisms. In \S\ref{sec:patheffect} we
present a decomposition that uses information interventions to achieve
a path-specific decomposition of agency causal effects, specifically
the average treatment effect. In \S\ref{sec:estima} we develop
non-parametric efficient estimators for some of the causal influence
measures proposed, and in \S\ref{sec:aplica} we present the results of
applying the methods to two publicly available datasets.  We conclude
in \S\ref{sec:discussion} with a brief discussion of connections to recent
literature and directions of future research.

\section{Notation, data, and causal model}\label{sec:notation}
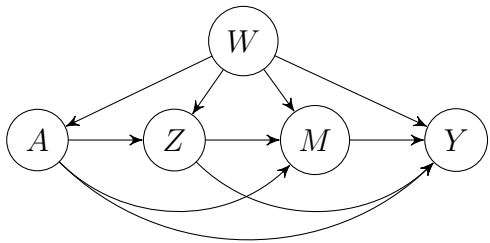
\begin{figure}[!htb]
  \centering
  \begin{tikzpicture}
    \tikzset{line width=1pt, outer sep=0pt,
      ell/.style={draw,fill=white, inner sep=2pt,
        line width=1pt}};
    \node[circle, draw, name=a, ]{$A$};
    \node[circle, draw, name=z, right = 10mm of a]{$Z$};
    \node[circle, draw, name=m, right = 10mm of z]{$M$};
    \node[circle, draw, name=y, right = 10mm of m]{$Y$};
    \node[circle, draw, name=w, above right = 7mm and 3mm of z]{$W$};

    \draw[->](a) to (z);
    \draw[->](w) to (z);
    \draw[->](z) to (m);
    \draw[->](a) to[out=-45,in=225] (y);
    \draw[->](a) to[out=-45,in=225] (m);
    \draw[->](z) to[out=-45,in=225] (y);
    \draw[->](w) to (y);
    \draw[->](m) to (y);
    \draw[->](w) to (a);
    \draw[->](w) to (m);
  \end{tikzpicture}
  \caption{Causal graph associated to the SCM in equation
    (\ref{eq:npsem}), where the interest is in measuring the causal
    relation between $A$ and $Y$ and quantifying how it operates
    through $M$. For illustrative convenience the exogenous variables
    $U$.}
  \label{fig:dag}  
\end{figure}

Assume that we observe $n$ independent and identically distributed
copies $X_1,\ldots,X_n$ of
$X\allowbreak=(W,A,Z,\allowbreak M,Y)\sim \P$. We use a Structural
Causal Model \citep[SCM, ][]{Pearl00} to study causal relations. A SCM
is a 4-tuple $\langle U, X, \mathcal F, \mathcal P\rangle$, where $U$
are exogenous variables, $X$ are endogenous variables, $\mathcal F$ is
a set of functions used to determine the value of $X$, and
$\mathcal P$ is a set of allowed distributions on $U$
\citep{bareinboim2022pearl}. The SCM used in this paper assumes the following.
\begin{assumptioniden}\label{ass:PU}
  Assume that $X$ is generated according to:
  \begin{equation}\label{eq:npsem}
    \begin{gathered}
      W=f_W(U_W);\qquad
      A=f_A(W,U_A);\qquad
      Z=f_Z(A,W,U_A);\\
      M=f_M(Z,A,W,U_M);\qquad
      Y=f_Y(M,Z,A,W,U_Y).
    \end{gathered}
  \end{equation}
  where the functions $f$ are deterministic but otherwise
  unrestricted. Furthermore, assume
  $U_A \indep (U_Y,U_M,\allowbreak U_Z)\mid W$,
  $U_Z\indep (U_Y,U_M)\mid (A,W)$, and $U_M\indep U_Y\mid (A,W,Z)$.
\end{assumptioniden}
We simplify the presentation of the model by considering a single $W$
with little loss of generality, but we could split it into several
factors according to whether they are confounders of only some of the
subsequent relations.     We are interested in quantifying the strength of the causal
relation between $A$ and $Y$, and understanding the extent to
which that relation operates through the various paths between the
endogenous variables $X$. We illustrate those pathways in the
causal graph depicted in Figure~\ref{fig:dag}. For illustrative
convenience we do not depict the exogenous variables; correlations
between them could be depicted through bi-directed arrows (see
\cite{Pearl00} for the construction of a causal graph from an
SCM).
Variables such as $Z$, which are affected by $A$ and are a common
cause of the mediator and outcome, are often referred to as
intermediate confounders. In this causal graph, one may be interested in
mediation analysis in the sense of quantifying all paths from $A$ to
$Y$ that involve $M$ (indirect paths) vs all other paths (direct
paths). An alternative and arguably more complete analysis would
involve understanding the causal relation between $A$ and $Y$ as it
operates through each path $A\to Y$, $A\to Z\to Y$, $A\to Z\to M\to
Y$, and $A\to M\to Y$. We refer to the former goal as
\textit{mediation analysis}, and to the latter as \textit{path-analysis}.

\section{Measuring the strength of causal relations}\label{sec:total}

The concept of \textit{causal influence} is generally defined as the
existence of a directed path between two variables
\citep{pearl1995theory}. A formal definition of the causal influence
of $A$ on $Y$ in terms of the non-parametric structural equation model
can be constructed as follows.
\begin{definition}[Causal influence]\label{def:infl}
  For fixed $a$, let the counterfactual variable $Y(a)$ be defined as
  the solution in $Y$ of the equations (\ref{eq:npsem}) under the
  intervention $A=a$. That is, $Y(a)=f_Y(M(a),Z(a),a,W,U_Y)$ with
  $M(a)=f_M(Z(a), a, W, U_M)$ and $Z(a)=f_Z(a, W, U_Z)$. The variable
  $A$ is said to have a \textit{causal influence} on $Y$ if and only
  if $\sup_{a,a'}|Y(a) - Y(a')|>0$ with positive probability.
\end{definition}  
For binary treatments, the concept of causal influence as stated above
is related to Fisher's sharp null hypothesis of no individual
treatment effect $H_0:Y_i(1)=Y_i(0)$ for all $i\in\{1,\ldots,n\}$. 
Recent approaches to causal inference have focused on quantifying
causal influence through causal effects, which are usually understood
as changes in the probability distribution of $X$ under hypothetical
interventions that modify the value of the variable $A$. In this
paper, we propose an alternative way to quantify causal influence,
using an approach based on intervening on the information transferred
along edges in the causal graph depicted in Figure~\ref{fig:dag}. In general,
we will require that measures for causality satisfy the following
properties:
\begin{property}[Sharp null criterion]\label{prop:sharp}
  A causality measure satisfies the sharp null criterion if it is null
  whenever there is no causal influence.
\end{property}
\begin{property}[Monotonicity criterion]\label{prop:mono}
  A causality measure satisfies the monotonicity criterion if (i) it is
  less than its null value whenever $Y(a)$ is monotone decreasing in
  $a$ almost surely, and (ii) it is greater than its null value
  whenever $Y(a)$ is monotone increasing in $a$ almost surely.
\end{property}

While this criterion is trivially satisfied by many causal effects, it
has been recently uncovered that it is not satisfied by widely adopted
approaches to mediation analysis. In \S\ref{sec:med} we prove that our
proposed path-specific causal influence measures satisfy a version of
this criterion adapted to path-analysis. Before we describe our
proposed definitions of intervention on information transferred along
edges, we review causal effects defined with respect to interventions
on nodes.

\subsection{Causal effects}\label{sec:ate}

A general definition of causal effect based on node interventions is
as follows \citep[adapted from][]{Pearl00}:
\hlrev{\begin{definition}[Causal effect of a node intervention] Given
    a fixed value $a$, the causal effect of the intervention $A=a$ is
    defined as a comparison between the probability distributions of
    the counterfactual variables $Y(a)$ and $Y(a_0)$, for a reference
    level $a_0$.
\end{definition}}
As stated in the definition, to define the causal effect of $A$ on $Y$
one needs to specify an intervention to the node corresponding to $A$
in the SCM (\ref{eq:npsem}). For example, consider an intervention
that removes the equation $f_A$ from the SCM and replaces it with
assignment to a fixed value $A=a$.  The counterfactual outcome in this
hypothetical intervened world is $Y(a)=f_Y(M(a),Z(a),a,W,U_Y)$,
denoting outcome that would have been observed in a world where,
contrary to fact, $A$ had been equal to $a$ with probability one. The
causal effect of $A$ on $Y$ may then be defined in terms of contrasts
of the distribution of the potential outcomes $Y(a):a\in\mathcal
A$. For example, for binary $A$, a common choice is to use the
\textit{average treatment effect} $\ate=\E[Y(1)-Y(0)]$ as an effect
measure.

The above definition has been generalized in multiple directions, for
example to allow dynamic interventions that might depend on $W$
\citep[e.g.,][]{robins2004optimal, van2005history, cain2010start},
modified treatment policies whereby the intervention might depend on
both $A$ and $W$ \citep[e.g.,
][]{diaz2020causal,diaz2021nonparametric, diaz2022causal}, as well as
stochastic interventions whereby the post-intervention exposure is a
random draw from a user-given distribution \citep[e.g.,
][]{wen2021intervention}. 


\subsection{Causal influence based on information
  interventions}\label{sec:strength}

\subsubsection{A motivating experimental design}

To motivate the definitions that follow, we fist discuss \textit{audit
  studies}, which are a type of experiment to study discrimination
common in social sciences \citep{gaddis2018audit}. Although there are
many types of audit studies, on a basic level they can be described as
a research design where the experimenter prompts some type of
interaction with the research subjects (e.g., a letter, a job or
housing application, a diagnosis, etc.), randomly varying the
attributes under evaluation (e.g., gender, race), and then assesses
the behavior of the research subject. To ground ideas, consider the
following examples.

\cite{shapiro2018implicit} study implicit biases in periviability
counseling among neonatologists caring for pregnant women by using
implicit association tests \citep[IATs][]{greenwald1998measuring}. The
experiment consisted of showing different neonatologists a vignette of
a woman in imminent labor, where the patient's race and socioeconomic
status in the vignette were randomized. The neonatologists were then
asked to indicate how likely they were to recommend comfort care vs
intensive care, and the difference in recommendation across randomized
races were assessed. \cite{bertrand2004emily} performed a similar
experiment to study the effect of perceived race on labor market
discrimination, where they randomized racial cues such in the
candidate's resume such as the name. \cite{cheng2020value} assess the
value of study abroad experience in the labor market. Their experiment
consisted of submitting resumes to job listings posted on a popular
online job site, where other resume attributes are held constant but
study abroad experience is randomized within job posting.

A key feature of all of these examples is that the intervention is
performed on the information given to a decision-maker (clinicians,
recruiters, etc.) rather than the causes (e.g., a person's race,
socioeconomic status, study abroad experience), which are left
unmodified. The definitions that follow can be seen as a formalization
of the interventions intended by these audit studies. 


\subsubsection{Proposed interventions on information transferred along
edges}

We propose to measure the causal influence of $A$ on $Y$ based on the
impact on the joint distribution of $(A,Y)$ of intervening on the
information transferred along all directed paths from $A$ to $Y$ in
Figure~\ref{fig:dag}. We first define an intervention where no
information is transferred along any of these paths. Specifically, let
$\A$ denote a noise variable defined as a random draw from the
distribution of $A$ conditional on $W$. Note that in some applications
it may be more reasonable to define $\A$ as a random draw from the
conditional distribution of $A$ given a subset of $V\subseteq W$. In
the interest of simplifying notation we pursue a definition based on a
random draw conditional on $W$.

To remove the information transferred along all paths between $A$ and
$Y$, one possibility is to transfer noise along the edges in
$S=\{A\to Y, A\to M, A\to Z\}$.
which corresponds to the following data generating mechanism:
  \begin{equation}\label{eq:npsem_mod}
    \begin{gathered}
    W=f_W(U_W);\qquad
    A=f_A(W,U_A);\qquad
    Z(\A)=f_Z(\A,W,U_A);\\
    M(\A)=f_M(Z(\A),\A,W,U_M);\qquad
    Y(\A)=f_Y(M(\A),Z(\A),\A,W,U_Y).
    \end{gathered}
  \end{equation}
  
  The above counterfactual represents interventions that remove the
  influence of $A$ on $Y$ along all edges in the set $S$ and replace it
  by transferring noise. We will therefore use the alternative
  notation $Y_S$ for $Y(\A)$. This notation will be useful when we
  define counterfactuals that intervene on the information transferred
  along different sets of edges, such as those required for path
  analysis.

  \begin{remark}
    The above interventions are stochastic versions of the
    \textit{info-interventions} proposed by \cite{heyang2019info}.
    These interventions are similar to the inflation operation
    proposed by \cite{wolfe2019inflation}, but are different in at
    least two aspects. First, our goal is to quantify the effect of a
    given cause on a given outcome, whereas the goal of inflation
    operations is to assess whether a particular data distribution is
    compatible with a given graphical model. Furthermore, our
    interventions are performed at the individual level, whereas the
    inflation operation is performed at the population level by
    altering the probability distribution of the data.
\end{remark}

These interventions may be of interest when, instead of considering
interventions on an individual's $A$, we consider a hypothetical joint
intervention on the mechanisms downstream of $A$ such that the
functional dependence on of these variables on their causal parents
are modified. For example, $Z = f_Z(A, W, U_{Z})$ is replaced by
$Z = f_Z(\A, W, U_{Z})$. The form of the function $f_Z$ remains the
same but one of its inputs changes. To further illustrate the
relevance of this definition, consider a problem where one is
interested in the causal influence of gender $A$ on wages $Y$,
mediated by work sector $M$ and field of study $Z$. The correlation
between gender $A$ and the counterfactual $Z(\A)$ is interpreted as
the correlation between gender and the field of study that would be
observed in a hypothetical world where field of study preferences and
opportunities were as if a person's actual gender did not play a role
in their field of study preferences and opportunities.

This information transfer intervention allows us to define causal
influence measures as follows:

\begin{definition}[Measure of causal influence]\label{def:strength}
  Let $\P(y,a)$ denote the distribution of $Y$ conditional on $A$, and
  let $\P_S(y,a)$ denote the distribution of $Y_S$ conditional on
  $A$. Let $D$ denote any functional contrasting two distributions
  such that $D(\P,\P)=0$.  The strength of the causal influence of $A$
  on $Y$ can be defined as any contrast $D(\P, \P_S)$ between (moments
  of) $\P(y,a)$ and $\P_S(y, a)$.
\end{definition}

\begin{remark}
  This definition is related to the definition of causal strength
  given by \cite{sprenger2018foundations,
    fitelson2011probabilistic}. The difference with our definition is
  that we are interested in a system where the information transferred
  along by the causal agent is changed to be noise, whereas the
  definition in these works focuses on contrasts where the causal
  agent is turned on vs off.
\end{remark}

To further see why Definition~\ref{def:strength} is a sensible measure
of causal influence, consider the original SCM (\ref{eq:npsem}) and its
associated causal graph in Figure~\ref{fig:dag}. In this model $A$ and
$Y$ may be associated due to two different types of relations: (i) any
of the directed paths from $A$ to $Y$, or (ii) undirected paths
operating through common causes (e.g., $W$ or $U$). In contrast, in
the intervened SCM (\ref{eq:npsem_mod}), $A$ and $Y_{S}$ can only be
associated due paths operating through to common causes.  Therefore, a
contrast between the joint distributions $\P(y, a)$ and $\P_S(y, a)$
provides a measure of the information transferred through paths from
$A$ to $Y$. These ideas are formalized in the following property:

\begin{proposition}
  Any contrast $D(\P, \P_S)$ between (moments of) $\P(y, a)$ and $\P_S(y, a)$
  satisfies the sharp null criterion \ref{prop:sharp}.
\end{proposition}
\begin{proof}
  This follows after noticing that $Y_{S}=Y$ whenever $A$ has no
  causal influence on $Y$.
\end{proof}

Note, however, that the counterfactual SCM (\ref{eq:npsem_mod}) should
not be used in general to define individual causal effects, since $\A$
may be equal to $A$ by chance for any given individual. Furthermore,
contrasting the marginal distributions of $Y_S$ and $Y$ may also be
misleading, as these distributions may also be equal by study design,
e.g., in a completely randomized trial with binary treatment.

To properly place the definition of the measure of causal influence
$D(\P, \P_S)$ in the context of causality research, it is useful to
consider Judea Pearl's ``hierarchy (or ladder) of causation''
\citep{PearlMackenzie18}. The hierarchy consists of three layers:
associational, agential,\footnote{The agential layer is often referred
  to as ``interventional'', although this is possibly a misnomer since
  the counterfactual layer also uses interventions
  \citep{Woodward2003}.} and counterfactual, which correspond to ``the
ordinary human activities of seeing, doing, and imagining,
respectively.'' \citep{bareinboim2022pearl}. In this hierarchy, the
distribution $\P_S(a,y)$ may be seen as that of the outcome and cause
in an imagined world where the cause under evaluation $Y$ does not
causally affect any of downstream variables $(Z,M,Y)$. Although we
operationalize this imagined world through the counterfactual
distribution $\P_S(a,y)$ and associated counterfactual variables,
there are a multiple other ways in which this imagined world could
have been operationalized.

To illustrate the utility of Definition~\ref{def:strength}, we now
present a number of examples of how these distributions can be
contrasted.
\begin{example}[Covariance decomposition]\label{ex:covar}
  Consider
  the covariance decomposition
  \[\cov(A,Y) = \underbrace{\cov(A,Y-Y_{S})}_{\text{causal influence}}
    +\underbrace{\cov(A,Y_{S})}_{\text{confounding}}.\] By definition,
  $A$ does not have a causal influence on $Y_{S}$. Any association
  between $A$ and $Y_{S}$ is due to common causes, which means that
  $\cov(A, Y_{S})$ is a measure of confounding. On the other hand,
  $\cov(A,Y-Y_{S}) = \cov(A,Y) - \cov(A,Y_{S})$ is a contrast of
  covariances comparing hypothetical worlds that only differ in the
  influence through paths from $A$ to $Y$, which is present in
  $\cov(A,Y)$ but not in $\cov(A,Y_{S})$. Thus, $\cov(A,Y-Y_{S})$
  quantifies the causal influence of $A$ on $Y$. Furthermore, it
  recovers the sign of the individual effects in the following sense:
  \begin{theorem}\label{theo:mono}
    The above covariance decomposition satisfies the monotonicity
    criterion \ref{prop:mono}.
  \end{theorem}

\end{example}

\begin{example}[Regression of residuals]\label{ex:residuals}
  Consider the function $f(a) = \E[Y-Y_{S}\mid A=a]$, where we note
  that this parameter measures the strength of the association in a
  hypothetical world where the influence of $A$ on $Y$ is removed (and therefore
  all association is due to confounding) and compares it to the
  association observed in the actual world.
\end{example}

\begin{example}[Kullback-Leibler divergence]\label{ex:kullback}
  Another measure of causal influence may be given by the
  Kullback-Leibler divergence
  $D(\P,\P_S)=\E_\P\left[\log\left(\frac{\p(Y,A)}{\p_S(Y,A)}\right)\right]$,
  where $\p(y,a)$ is the density of $(Y,A)$ and $\p_S(y,a)$ is the
  density of $(Y_{S},A)$.
\end{example}

\begin{remark}
  \cite{janzing2013quantifying} also propose a measure of causal
  influence based on the Kullback-Leibler divergence, but their proposal
  differs from ours in two ways. First, our interventional densities
  allow for random draws $\A$ conditional on the past, theirs do
  not. Second, their measure of causal influence is based on a
  contrast of the distribution of the complete data $X$ with and
  without an intervention. While using the distribution of the
  complete data $X$ yields causal influence measures with important
  properties \citep[see properties P0-P4
  in][]{janzing2013quantifying}, adopting a definition based on the
  distribution of $(A,Y)$ allows the construction of optimal
  estimators which would not be possible under the definition that
  contrasts the full data distributions.
\end{remark}

The parameters described in Examples~\ref{ex:covar}-\ref{ex:kullback}
have an interesting structural interpretation in terms of a
quantification of the dependence of the function $f_Y$ in the SCM on
the variable $A$. In other words, these parameters quantify how much
nature's causal mechanisms use information on $A$ to assign the
outcome value. The following theorem provides an expression that can
be used to identify the joint distribution of $(Y_{S}, A)$ under the
standard assumption of no unmeasured confounders.

\begin{theorem}[Identification]\label{theo:total}
  Under the assumed SCM (\ref{ass:PU}) we have
  $\P_S(Y_{S}\leq y\mid A=a) =\E[\P(Y\leq y\mid W)\mid A=a]$.
\end{theorem}
Note that identification of this causal influence parameter does not
require the positivity assumption $\P(\g(a\mid W))>0$ that is required
to assess the effect of many node interventions. Furthermore, if $A$
is randomized, this identification result reduces to
$\P(Y\leq y\mid A=a)$, in agreement with the idea that the causal
relation between $A$ and $Y$ is not confounded in a randomized
experiment.

\begin{coro} Under the assumed SCM (\ref{ass:PU}) we have
  $\cov(A,Y-Y_{S})=\theta$ and $\cov(A,Y_{S})=\tau$, where
  $\theta=\E\left\{\left[A - \E(A\mid W)\right]\left[Y - \E(Y\mid
      W)\right]\right\}$, and
  $\tau = \cov\left\{\E(A\mid W), \E(Y\mid W)\right\}$.
  Furthermore, we have $f(a) = \E[Y-\E(Y\mid W)\mid A=a]$.
\end{coro}

\begin{remark}
  \hlrev{The expectation of the conditional covariance $\theta$ has
    been previously used for causal inference as a means to study
    other causal effects, such as the variance-weighted ATE
    \citep{li2011higher}. Furthermore, it forms the basis to construct
    the partial correlation coefficient, which has a long but
    non-rigorous history as a measure of causality in applications
    \citep{ellett1986correlation}, for example in the context of
    genomics \citep[e.g., ][]{freudenberg2009partial}. Its role in
    conditional independence testing has been previously studied by
    \cite{10.1214/19-AOS1857}. However, we know of no previous result
    that provides an interpretation of the expected conditional
    covariance as a causal effect in terms of formal interventions in
    a causal model. In addition, this result provides a causal
    interpretation of the well known law of total covariance as a
    decomposition of the covariance between $A$ and $Y$ in terms of a
    pure causal influence $\theta$ and a pure confounding effect
    $\tau$.  Furthermore, identification of the contrast
    $\E[Y-Y_{S}\mid A=a]$ involves regressing the orthogonalized
    outcome $Y-\E(Y\mid W)$ on the treatment variable $A$. This is a
    procedure commonly used in applied studies aiming to estimate
    causal inference effects (e.g., in genomics) 
  but its interpretation in a formal causal inference framework has
  not been previously articulated.  This illustrates that causal our
  proposed causal influence measures formalize vague notions of causality that
  are already present in scientists minds.}
\end{remark}

The counterfactual $Y_S$ involves information transfer interventions
that remove all the information transferred along edges in $S$. In
this article we will also use an information transfer operation that
emulates information transferred along certain paths. Consider, for
example, the paths $A\to Z\to Y$ and $A\to Z\to M\to Y$. Let $Z_A$
denote a draw from the distribution of $Z$ conditional on $(A,W)$, and
consider the counterfactual variable $Y(A, Z_A, M(A, Z_A))$. In this
counterfactual variable, we have emulated the information transferred
along the paths $A\to Z\to Y$ and $A\to Z\to M\to Y$ by information
transferred along synthetic paths $A\to Z_A\to Y$ and
$A\to Z_A\to M\to Y$, respectively. Interestingly, the distribution of
this counterfactual variable is identified by $\P(Y\leq y\mid A=a)$,
suggesting that this information transfer intervention preserves the
relation that operates through the original paths. While this kind of
intervention is not very useful to define total causal influence of
$A$ on $Y$, it will be fundamental in \S\ref{sec:med} when we present
the path analysis methods using causal influence.

In the next section we discuss a major advantage of the information
transfer interventions introduced in this section, namely the ability
to provide measures of direct and indirect causal influence. We will
show that these interventions allow the decomposition of the total
causal influence into an influence and an effect that operates through
each specific path. Importantly, we show that this decomposition is
possible even in the presence of mediator-outcome confounders which
are affected by exposure, a problem whose solution has been elusive in
the causal inference literature that focuses on the causal effect of
actions. We start our presentation with a brief discussion of the
problems with existing approaches to mediation analysis using causal
effects.

\section{Causal effects and implications for
  path and mediation analysis}\label{sec:med}

Mediation analysis is the task of decomposing the total causal
influence of $A$ on $Y$ into an influence that operates through $M$
and an influence that operates through all other mechanisms. This can
be done by decomposing the total causal influence of the previous
section into influence that operates the pathways $A\to Y$ and
$A\to Z\to Y$ (so-called direct influence), and influence that
operates through the pathways $A\to M\to Y$ and $A\to Z\to M\to Y$
(so-called indirect influence).

\begin{definition}[Causal influence through a
  mediator]\label{def:infl_med}
  For fixed $\bar a = (a_0, a_1, a_2)$, define the counterfactual
  variable $Y(a_0, M(a_1))$ as the solution in $Y$ of the equations
  (\ref{eq:npsem}) under the intervention $A=a_0$ and $M=M(a_1)$. That
  is, $Y(a_0,M(a_1))=f_Y(M(a_1),Z(a_0),a_0,W,U_Y)$ with
  $M(a_1)=f_M(Z(a_1), a_1, W, U_M)$ and $Z(a_0)=f_Z(a_0, W, U_Z)$. The
  variable $A$ is said to have a causal influence on $Y$ through $M$
  if and only if $\sup_{\bar a}|Y(a_0, M(a_1)) - Y(a_0, M(a_2))|>0$
  holds with positive probability.
\end{definition}  

For binary exposures, this definition of causal influence is
equivalent to the sharp mediation null $H_0: Y(a,M(1)) = Y(a,M(0))$
for $a\in\{0,1\}$. As before, it will be desirable that parameters
that measure the influence of $A$ on $Y$ operating through $M$ satisfy
the following properties \citep{miles2022causal}:

\begin{property}[Mediational sharp null criterion]\label{prop:medsharp}
  A measure of the causal influence of $A$ on $Y$ through $M$ is said
  to satisfy the mediational sharp null criterion if it is null
  whenever there is no causal influence through the mediator $M$.
\end{property}


In addition, it may be desirable that mediation analyses that seek to
unveil mechanisms satisfy the following property, which we present in
an additive scale but can be equivalently stated in a multiplicative
or other scales:

\begin{property}[Total influence decomposition]\label{prop:totdecomp}
  A pair of measurements of direct and indirect influence are said to
  decompose a measure of total influence if they add up to the total
  influence.
\end{property}

Having established the above desiderata for a measure of influence
through a mediator, we now review two of the major mediation
frameworks recently proposed: natural direct and indirect effects, and
randomized interventional direct and indirect effects. We discuss the
lack of identifiability of natural effects in the presence of a
mediator-outcome confounder affected by treatment, and discuss the
fact that randomized mediational effects do not satisfy the sharp
mediational null criterion. These shortcomings were originally
described by \cite{avin2005identifiability} and
\cite{miles2022causal}, respectively. We then move onto discussing our
proposal for mediation analysis based on the causal influence measures
proposed in the previous section, and show how our proposal can be
used to address those shortcomings.

\subsection{Natural direct and indirect effects}

In the case of a binary exposure $A$, the average treatment effect
$\E(Y(1)-Y(0))$ is a common measure of the effect of the action $A=1$ vs
the action $A=0$. Natural mediation effects decompose the ATE into
effects that operate through the mediator $M$ and effects that operate
through all other causes. Specifically, we have
\begin{align*}  
  \E[Y(1)-Y(0)] &= \E[Y(1,M(1))-Y(0,M(0))] \\&=
  \underbrace{\E[Y(1,M(1))-Y(1,M(0))]}_{\text{Natural indirect effect
  (NIE)}} + \underbrace{\E[Y(1,M(0)) - Y(0,M(0)))}_{\text{Natural
  direct effect (NDE)}},
\end{align*}
where the first equality follows by definition of
$Y(a,M(a)):a\in\{0,1\}$ and the second by adding and subtracting
$\E[Y(1,M(0))]$. Inspection of the above formulas reveal why the NIE
measures the effect through $M$ and the NDE measures the effect
through all other mechanisms. The NIE is the result of comparing the
outcome in a hypothetical world where $A=1$ is fixed while $M$ is
varied from what it would have been under $A=1$ to what it would have
been under $A=0$. The NDE is the result of comparing the outcome in a
hypothetical world where $A$ is varied from $A=1$ to $A=0$ while
fixing $M$ to what it would have been under $A=0$.

While the NIE and NDE provide a useful and intuitive decomposition of
the ATE into effects that operate through $M$ vs all other mechanisms,
and satisfy \ref{prop:medsharp} and \ref{prop:totdecomp}, these
effects are not identified in the causal graph~\ref{fig:dag}
\citep{avin2005identifiability}. To understand why identification
fails, it is useful to consider a simplified model where
$W=\emptyset$, all the errors $U$ are mutually independent, and the
edges $A\to M$ and $A\to Y$ have been removed. It can be proved that
in this model we have
{\small
\[\E[Y(1,M(0))]=\sum_{m,z,z'} \E[Y(Z=z,M=m)]\P(M(Z=z')=m)\P(Z(A=1)=z, Z(A=0)=z'),\]}%
where we have enriched the notation to add the intervention node in
the index of the counterfactual variables. While $\P(Y(Z=z,M=m)=y)$
and $\P(M(Z=z')=m)$ are identifiable in this model,
$\P(Z(A=1)=z, Z(A=0)=z')$ is not \citep[for a counterexample, see
Table 1 of][]{avin2005identifiability}, leading to lack of
identifiability of the NIE and NDE. The variable $Z$ is often referred
to as a \textit{recanting witness} because it operates as an direct
mechanism through the path $A\to Z\to Y$ and as an indirect mechanism
through the path $A\to Z\to M\to Y$, leading to the lack of
identifiability of either the direct or the indirect effect.

Multiple alternatives exist to conduct mediation analysis in the
presence of intermediate confounding. For example, if the relation
$A\to Z$ is monotonic, \cite{tchetgen2014identification} showed that
the NDE and NIE are point-identified and \cite{rudolph2023efficient}
developed non-parametric estimators for the identifying
functionals. Alternatively, \cite{miles2015partial} gave partial
identification bounds for the NDE and NIE. A third option is to target
a different definition of direct and indirect effects, for example
using the methods. Along these lines, much of the recent literature on
mediation analysis has focused on \textit{randomized interventional
  effects} \citep{didelez2006direct, van2008direct,
  vanderweele2014effect, diaz2021nonparametricmed}. We discuss the
definition, interpretation, and identification of these randomized
interventional effects below.

\subsection{Randomized interventional direct and indirect effects}

Consider a random draw $G(a)$ from the distribution of $M(a)$
conditional on $W$. Randomized mediational effects are concerned with
interventions that set the mediator to $G(1)$ and $G(0)$ instead of
$M(1)$ and $M(0)$. Specifically, the randomized mediational effects are
defined as follows:
\[\underbrace{\E[Y(1,G(1))-Y(0,G(0))]}_{\text{Total effect}} = 
  \underbrace{\E[Y(1,G(1))-Y(1,G(0))]}_{\text{Randomized indirect
      effect}} + \underbrace{\E[Y(1,G(0)) -
    Y(0,G(0))]}_{\text{Randomized direct effect}}.\] The first
limitation of randomized mediational effects is that they do not
satisfy \ref{prop:totdecomp} in the sense that they do not decompose
the ATE, but rather decompose an alternative treatment effect given by
the left hand side of the above expression which is defined in terms
of interventions on both $A$ and $M$.


Unlike the NIE and NDE, randomized mediational effects are identified
in the model in the SCM~(\ref{eq:npsem}). \cite{vanderweele2014effect}
show that the above randomized interventional indirect effect is
identified as
{\small
\begin{equation}\E[Y(a,G(a'))]=\int \E(Y\mid
  m,z,a,w)\dd\P(z\mid a,w)\dd\P(m\mid a',w)\dd\P(w)\label{eq:idenrie}
\end{equation}}%
under the assumption that there are no
unmeasured confounders of the relations $A\to M$, $A\to Y$, and
$M\to Y$. However, \cite{miles2022causal} has recently uncovered an
important limitation of these effects, namely that they fail to
satisfy the mediational sharp null criterion \ref{prop:medsharp}. One
counterexample involves creating an SCM with independent errors,
$W=\emptyset$, and an exogenous binary variable $U_Z$ such that
$M(a) = M(a')$ in the event $U_Z=1$ and $Y(a,m') = Y(a,m)$ in the event
$U_Z=0$ for all $a$, $a'$, and $m$, such that there is no causal
influence of $A$ on $Y$ operating through $M$ (Definition
\ref{def:infl}). It is easy to see that the randomized interventional
indirect effect  in this example
equals
\[\int\E(Y\mid m, z,a)\dd\P(z\mid a)[\dd\P(m\mid a)-\dd\P(m\mid a')],\]
which is generally not null, and would only be null if $U_Z$ was observed
and conditioned upon in all the above quantities. The interested reader is referred to
\cite{miles2022causal} for more details and counterexamples. 

In what follows we propose a mediation analysis strategies based on
the information transfer interventions introduced in
\S\ref{sec:total}, and show that this approach overcomes the
limitations of the interventional effects discussed in this section.

\section{Path analysis using causal influence}\label{sec:pathanal}

In this section, we propose a decomposition of the causal influence of
$A$ on $Y$ into influence that operates through each of the paths
$A\to Y$, $A\to Z\to Y$, $A\to Z\to M\to Y$, and $A\to M\to Y$. We
first state the properties that are desirable of measures of such
path-specific influence. In this section, we will use counterfactual
variables indexed by interventions on $(A,Z,M)$, and define
$Y(a,z,m)=f_Y(m,z,a,W,U_Y)$, $M(a,z)=f_M(z,a,W,U_M)$, and
$Z(a)=f_Z(a, W, U_Z)$.

\begin{definition}[Path-specific causal influence]\label{def:infl_path}
  For fixed $\bar a = (a_0, a_1, a_2, a_3, a_4)$, define the
  counterfactual variable $Y(a_1, Z(a_2), M(a_3, Z(a_4)))$. The
  variable $A$ is said to have a causal influence on $Y$ through each
  path $P_1$, $P_2$, $P_3$, $P_4$ if and only if the following
  conditions hold with positive probability \renewcommand{\arraystretch}{1.5}
  \begin{table}[H]
    \centering
    {\small
      \begin{tabular}[H]{c|l|c}\hline
        Name &  Path & Condition \\ \hline
        $P_1$ & $A\to Y$ &  $\sup_{\bar a}|Y(a_1, Z(a_2), M(a_4, Z(a_3))) -
                           Y(a_0, Z(a_2), M(a_4, Z(a_3)))|>0$\\        
        $P_2$ &  $A\to Z\to Y$ &  $\sup_{\bar a}|Y(a_1, Z(a_2), M(a_4, Z(a_3))) -
                                 Y(a_1, Z(a_0), M(a_4, Z(a_3)))|>0$\\
        $P_3$ &  $A\to Z\to M\to Y$ &  $\sup_{\bar a}|Y(a_1, Z(a_2), M(a_4, Z(a_3))) -
                                      Y(a_1, Z(a_2), M(a_4,
                                      Z(a_0)))|>0$\\
        $P_4$ &  $A\to M\to Y$ &  $\sup_{\bar a}|Y(a_1, Z(a_2), M(a_4, Z(a_3))) -
                                 Y(a_1, Z(a_2), M(a_0, Z(a_3)))|>0$\\
        \hline
      \end{tabular}
    }%
    \caption{Conditions for causal influence through each path.}
    \label{tab:conds}
  \end{table}
\end{definition}

\begin{property}[Path-specific sharp null criterion]\label{prop:sharp_ps}
  For $j=1,2,3,4$, a measure of the causal influence operating through
  path $P_j$ satisfies the path-specific sharp null criterion if it
  is null whenever there is no causal influence through $P_j$. 
\end{property}
In addition, it will be important that the measures of the causal
influence through each path recover the direction of the mechanism in
some sense. \cite{miles2022causal} operationalized this idea using the
following property:
\begin{property}[Path-specific monotonicity criterion]\label{prop:mono_ps}
  For $j=1,2,3,4$, a measure of the causal influence operating through
  path $P_j$ satisfies the monotonicity criterion if (i) it is less
  than its null value whenever $Y(a_1, Z(a_2), M(a_4, Z(a_3)))$ is
  decreasing in $a_j$ almost surely, and (ii) it is greater than its
  null value whenever $Y(a_1, Z(a_2), M(a_4, Z(a_3)))$ is increasing
  in $a_j$ almost surely, for all $a_s:s\neq j$.
\end{property}

Our proposed method for path analysis will require to specify a set of
interventions that sequentially remove the information transferred
through the paths $P_1$, $P_2$, $P_3$, and $P_4$. To construct the
interventions, we first define the sets $S_j = \{P_1,\ldots,P_j\}$ for
$j=1,2,3,4$. Then, let $\A$ and $Z_\A$ denote a random draw from the
distribution of $A$ and $Z$ conditional on $W$, respectively. Let
$Z_A$ denote a random draw from the distribution of $Z$ conditional on
$(A,W)$. The notation $Z_\A$ indicates that $Z_\A$ does not transfer
information from $A$ onto the descendants of $Z$, whereas the notation
$Z_A$ indicates that $Z_A$ transfers information from $A$ onto the
descendants of $Z$. In this sense, an intervention that assigns $Z$ as
$Z_A$ can be thought of as an \textit{edge-emulation}
intervention. Although the emulation of the information transferred
along the edge is not perfect (e.g., $Z_A$ cannot transfer information
from $U_Z$ into $M$), it will be sufficient for purposes of defining
path-specific causal influence. Define the following counterfactual
variables:
\begin{align*}
  Y_{S_0} & = Y(A, Z(A), M(A, Z(A))),     \\
  Y_{S_1} & = Y(\A, Z(A), M(A, Z(A))),    \\
  Y_{S_2} & = Y(\A, Z(\A), M(A, Z(A))),   \\
  Y_{S_3} & = Y(\A, Z(\A), M(A, Z(\A))),   \\
  Y_{S_4} & = Y(\A, Z(\A), M(\A, Z(\A))), 
\end{align*}
A straightforward measurement of the causal influence through path
$P_j$ could be achieved through a contrast of the distributions
$\P(Y_{S_j}\leq y\mid A=a)$ and $\P(Y_{S_{i-1}}\leq y\mid A=a)$, for
$j=1,2,3,4$. (A related definition of path-specific influence is
proposed by \cite{zhang2018non} in terms of a covariance contrast and
using atomic interventions.) However, some of these causal influence
measures are not identified due to the recanting witness problem
outlined in the previous section. Specifically, the probability
distribution of $Y_{S_2}$ is not identifiable because $Z$ operates as
a recanting witness through the paths $A \to Z\to Y$ and
$A \to Z\to M\to Y$. This means that the causal influence through the
path $A \to Z\to Y$, defined as a contrasts between
$\P(Y_{S_2}\leq y\mid A=a)$ and $\P(Y_{S_1}\leq y\mid A=a)$ is not
identified. The same is true for the causal influence through the path
$A \to Z\to M\to Y$, defined as a contrasts between
$\P(Y_{S_3}\leq y\mid A=a)$ and $\P(Y_{S_2}\leq y\mid A=a)$.

Fortunately, the edge emulation-intervention can be used to solve this
problem. Specifically, denote $Y_{S_j}$ with $Y_{S_j}^{(0)}$, and let
\begin{align}
  Y_{S_1}^{(1)} &= Y(\A, Z(A), M(A, Z_A)),   \notag\\
  Y_{S_2}^{(1)} &= Y(\A, Z(\A), M(A, Z_A)),   \label{eq:ctfs}\\
  Y_{S_2}^{(2)} &= Y(\A, Z_\A, M(A, Z(A))),   \notag\\
  Y_{S_3}^{(2)} &= Y(\A, Z_\A, M(A,Z(\A))).\notag
\end{align}
Then we have the following definition.

\begin{definition}[Measure of causal influence through a path]\label{def:strengthpath}
  For $i=0,1,2,3,4$ and $k=0,1,2$, let $\P_{S_j}^{(k)}(y,a)$ denote
  the distribution of $(Y_{S_j}^{(k)}, A)$. Let $D$ denote any
  functional contrasting two distributions.  The strength of the
  causal influence of $A$ on $Y$ operating through path $P_j$ can be
  defined as any contrast $D(\P_{S_{j-1}}^{(0)}, \P_{S_j}^{(0)})$ for paths
  \textit{not} involving the recanting witness $Z$ (i.e., $P_1$ and
  $P_4$), as $D(\P_{S_1}^{(1)}, \P_{S_2}^{(2)})$ for the path $P_2$, and as
  $D(\P_{S_2}^{(2)}, \P_{S_3}^{(2)})$ for $P_3$.
\end{definition}

\begin{remark}
  The above interventions are stochastic versions of those proposed in
  \cite{gong2021path}, but are also different in that we vary the
  information about $A$ transferred along a pathway, using different
  kinds of interventions depending on the pathway under
  measurement. This has important implications, as it allows us to
  prove important desiderata for path-analysis effect decompositions
  such as path-specific sharp null criteria
  (Theorem~\ref{theo:sharp}), as well as a result stating that the
  path-specific measures decompose the total effect
  (Theorem~\ref{theo:decomp}).
\end{remark}

In Theorem~\ref{theo:sharp} below we show that the above definition
satisfies the path-specific null criterion, meaning that these
parameters may be used to test the null hypothesis of no path-specific
effect. 

\begin{theorem}\label{theo:sharp}
  The contrasts defined in Definition~\ref{def:strengthpath} satisfy
  the path-specific sharp null criterion \ref{prop:sharp_ps} with
  respect to each path $P_j$.
\end{theorem}

Note that the definitions of the counterfactuals $Y_{S_1}^{(1)}$,
$Y_{S_2}^{(1)}$, $Y_{S_2}^{(2)}$, and $Y_{S_3}^{(2)}$, entail
intervening on the information that is transferred along certain
paths. For example, in $Y_{S_1}^{(1)}$ the influence of $A$ on $Y$
operating through the path $A\to Z\to M\to Y$ operates by means of the
random draw $Z_A$, whereas the influence operating through the path
$A\to Z\to Y$ operates through the natural value of $Z$. These edge
emulation interventions substitute a path $A\to Z\to\cdots\to Y$ by a
synthetic path $A\to Z_A\to\cdots\to Y$ that transfer the same
information as the original path. The edge-emulation intervention is
what allow us to do away with the recanting witness problem to obtain causal
influence measures that satisfy the path-specific sharp null
criterion.



The above definitions allow us to test the null hypothesis of no
causal influence through each path. A related goal is to decompose the
total influence of $A$ on $Y$ into influences operating through each
path. This is achievable only for some contrasts $D$ (e.g., it is not
achievable for the Kullback-Leibler divergence of
Example~\ref{ex:kullback}). Specifically, we have the following result
which is presented in an additive scale but could also be proved for a
multiplicative scale.

\begin{theorem}[Decomposition of the total influence into
  path-specific influences]\label{theo:decomp}
  Assume that the contrast function $D$ is linear in the sense that
  $D(\P, \F) = D(\P, \G) + D(\G, \F)$ for any distributions $\F$,
  $\P$, and $\G$.  Define the path-specific influences
  $\theta_{P_1} = D(\P_{S_0}^{(0)}, \P_{S_1}^{(0)})$,
  $\theta_{P_2} = D(\P_{S_1}^{(1)}, \P_{S_2}^{(1)})$,
  $\theta_{P_3} = D(\P_{S_2}^{(2)}, \P_{S_3}^{(2)})$, and
  $\theta_{P_4} = D(\P_{S_3}^{(0)}, \P_{S_4}^{(0)})$, as well as the
  parameter
  $\theta_{P_2\vee P_3} = D(\P_{S_1}^{(0)}, \P_{S_1}^{(1)}) + D(\P_{S_2}^{(1)},
    \P_{S_2}^{(2)}) + D(\P_{S_3}^{(2)}, \P_{S_3}^{(0)})$.
Then we have the following decomposition of the
total causal influence $\theta=D(\P_{S_0}, \P_{S_4})$:
\[\theta = \theta_{P_1} + \theta_{P_2} +  \theta_{P_3} +
  \theta_{P_4} +  \theta_{P_2\vee P_3}.\]
\end{theorem}

Clearly, the covariance contrast and the expectation contrast of
Examples~\ref{ex:covar} and \ref{ex:residuals} satisfy the assumption
of the theorem. In the above decomposition, the parameter
$\theta_{P_2\vee P_3}$ appears as a consequence of the addition of the
counterfactuals in Equation (\ref{eq:ctfs}). As the notation implies,
this parameter is equal to zero if there is no influence through the
path $P_2$, or there is no influence through the path $P_3$. This
result is proved formally in Proposition~\ref{prop:zero}
below. Intuition for this may be obtained as follows. The contrasts
$D(\P_1^{(0)},\P_2^{(0)})$ and $D(\P_2^{(0)},\P_3^{(0)})$, which would
readily yield measures of the influence through paths $P_2$ and $P_3$,
respectively, are not identified because the distribution $\P_2^{(0)}$
is not identified due to the recanting witness problem. However, the
contrast $\theta_{P_2\wedge P_3}=D(\P_1^{(0)},\P_3^{(0)})$ is
identified, and it measures the influence operating through $P_2$
\textit{and} $P_3$. That is, $\theta_{P_2\wedge P_3}=0$ if there is no
causal influence operating through $P_2$ nor $P_3$. This highlights
the difficulty in measuring the influence through paths $P_2$ and
$P_3$ separately. The above theorem decomposes
$\theta_{P_2\wedge P_3}$ into an influence measure $\theta_{P_2}$ that
operates only through $P_2$, an influence measure $\theta_{P_3}$ that
operates only through $P_3$, and an influence measure
$\theta_{P_2\vee P_3}$ that operates through $P_2$ \textit{or}
$P_3$. This latter parameter is zero if the influence going trough
\textit{either} of these paths is null. Furthermore, as demonstrated
in the following lemma, the parameter is null whenever there is no
intermediate confounding by $Z$.

\begin{proposition}\label{prop:zero}
  Assume the contrast $D$ is linear as in
  Theorem~\ref{theo:decomp}. In addition, assume that $D$ is such that
  $D(\P, \F) = \sum_{S\in\mathcal S} D(\P(\cdot\mid S), \F(\cdot\mid
  S))\P(S)$ for any partition $\mathcal S$ of the sample space, and
  that it satisfies $D(\F,\F)=0$. Let $\mathcal U$ denote the range of
  $U=(U_W, U_A, U_Z, U_M, U_Y)$. Assume $\mathcal U$ can be
  partitioned in sets $\mathcal U_1$, $\mathcal U_2$, and
  $\mathcal U_3$ such that the following hold almost surely
  \begin{itemize}
  \item $\sup_{\bar a}|Z(a_1) - Z(a_0)|=0$ in $\mathcal U_1$, and
  \item $\sup_{\bar z}|M(z_1) - Z(z_0)|=0$ in $\mathcal U_2$, and
  \item $\sup_{\bar z}|Y(z_1) - Y(z_0)|=0$ in $\mathcal U_3$.
  \end{itemize}Then $\theta_{P_2\vee P_3}=0$.
\end{proposition}

Importantly, the above proposition implies that whenever there is no
intermediate confounding, the path-specific decomposition is exact in
the sense that it satisfies property \ref{prop:totdecomp}. We now
illustrate the proposed path-analysis based on the covariance
decomposition of Example~\ref{ex:covar}.

\begin{example}[continues=ex:covar]
  Let $D$ denote the covariance contrast defined
  as $D(\P, \P_S) = \cov(A, Y - Y_S)$ for $(Y, A)\sim\P$ and
  $(Y_S,A)\sim \P_S$. Then, for $\theta=\cov(A, Y-Y_S)$, we have
  \[\theta = \theta_{P_1} + \theta_{P_2} +  \theta_{P_3} +
    \theta_{P_4} + \theta_{P_2\vee P_3},\]
  where
    $\theta_{P_1} = \cov(A, Y_{S_0}^{(0)} - Y_{S_1}^{(0)})$, 
    $\theta_{P_2} = \cov(A, Y_{S_1}^{(1)} - Y_{S_2}^{(1)})$, 
    $\theta_{P_3} = \cov(A, Y_{S_2}^{(2)} - Y_{S_3}^{(2)})$, 
    $\theta_{P_3} = \cov(A, Y_{S_3}^{(0)} - Y_{S_4}^{(0)})$,
and
\begin{equation}
  \theta_{P_2\vee P_3} =\cov(A, Y_{S_1}^{(0)} - Y_{S_1}^{(1)} +
  Y_{S_2}^{(1)}-Y_{S_2}^{(2)}+Y_{S_3}^{(2)}-Y_{S_3}^{(0)}).\label{eq:covdec}
\end{equation}
Furthermore, we have:
\begin{theorem}\label{theo:monopath}
  The covariance contrasts $\theta_{P_1}$, $\theta_{P_2}$,
  $\theta_{P_3}$, and $\theta_{P_4}$ satisfy the path-specific
  monotonicity criterion \ref{prop:mono_ps}.
\end{theorem}
\end{example}

\cite{wright1921correlation,wright1923theory,wright1934method}
proposed a covariance decomposition for $\cov(A, Y)$ in terms of
path-specific coefficients in the context of linear models and in the
absence of an intermediate confounder $Z$. An immediate consequence of
the above results is that our covariance decomposition generalizes
Wright's approach to a non-parametric model in the presence of
intermediate confounding. \cite{zhang2018non} provide an alternative
generalization that is unidentifiable in the presence of intermediate
confounders $Z$.


We now present identification formulas for the probability
distributions involved in computation of the measures of causal
influence of Definition~\ref{def:strengthpath}.

\begin{assumptioniden}[Overlap]\label{ass:overlap}
  For a fixed value $a$, assume the following hold for all $w$ such
  that $p(w)>0$:
  \begin{itemize}
  \item $\p(z\mid a,w)>0$ implies $\p(z\mid a',w)>0$
    for all $a'$ such that $\p(a'\mid w)>0$.
  \item $\p(m\mid z,a,w)>0$ implies $\p(m\mid z', a',w)>0$
    for all $(a',z',z)$ such that $\p(a'\mid w)>0$,
    $\p(z\mid a, w)>0$, and $\p(z'\mid a, w)>0$.
  \item $\p(m\mid z,a,w)>0$ implies $\p(m\mid z', a',w)>0$
    for all $(a',z',z)$ such that $\p(a'\mid w)>0$,
    $\p(z\mid a', w)>0$, and $\p(z'\mid a', w)>0$.
  \end{itemize}
\end{assumptioniden}

\begin{theorem}[Identification of path-specific causal influence]\label{theo:iden}
  Under the assumed SCM (\ref{ass:PU})
  plus \ref{ass:overlap}. the path-specific counterfactual
  distributions are identified as follows
  \begin{align*}
    \P(Y_{S_0}^{(0)}\leq y\mid A=a)& = \F(y \mid A=a),\\
    \P(Y_{S_1}^{(0)}\leq y\mid A=a)& = \E\left[\int \F(y\mid
                              a',Z, M,W)\dd\P(a'\mid W)\mid
                               A=a\right],\\
    \P(Y_{S_1}^{(1)}\leq y\mid A=a)& = \E\left[\int \F(y\mid
                               a',z, M,W)\dd\P(z\mid a, W)\dd\P(a'\mid
                                W)\mid A=a\right],\\
    \P(Y_{S_2}^{(1)}\leq y\mid A=a)& = \P(Y_{S_2}^{(2)}\leq y\mid A=a)\\
                             &=\E\left[\int \F(y\mid
                                a',z, M,W)\dd\P(z, a'\mid  W)\mid
                                A=a\right],\\
    \P(Y_{S_3}^{(2)}\leq y\mid A=a)& = \E\left[\int \F(y\mid
                                     a',z,m,W)\dd\P(m\mid a, z', 
                                     W)\dd\P(z\mid a',  W)\dd\P(z', a'\mid W)\mid
                                A=a\right],\\
    \P(Y_{S_3}^{(0)}\leq y\mid A=a)& = \E\left[\int \F(y\mid
                                a',z,m,W)\dd\P(m\mid a, z, 
                                     W)\dd\P(z, a'\mid  W)\mid
                                A=a\right],\\
    \P(Y_{S_4}^{(0)}\leq y\mid A=a)& = \E\left[\F(y\mid W)\mid A=a\right].
  \end{align*}
\end{theorem}

At this point it is important to note that the ideas of information
transfer interventions discussed in this paper can also be used to
obtain a decomposition of the average treatment effect into
path-specific effects analogous to the decomposition of
Theorem~\ref{theo:decomp}. We discuss such an extension in
\S\ref{sec:patheffect} of the supplement.

In the following section, we discuss efficient non-parametric
estimation of the covariance influence discussed in
Example~\ref{ex:covar}. Efficient non-parametric estimation of the
other parameters discussed in this paper is also possible, but we
defer the development of such estimators to future work.

\section{Extension to path analysis using causal effects}\label{sec:patheffect}

This section presents results where we use the information transfer
interventions introduced to construct a decomposition of the average
effect of a binary treatment into path-specific effects that satisfy
path-specific null criteria, analogous to the decomposition
constructed for measures of causal influence. A decomposition of the
ATE is important for questions that require an agency view of
causality, as the ATE can be used to quantify the effect of a
hypothetical but feasible intervention. The methods we present below
decompose the ATE, which can have an agency interpretation of
causality, into path-specific effects that do not correspond to an
agency interpretation of causality. Readers interested in
decompositions that have an agency interpretation are encouraged to
consult the literature on so-called separable effects \citep[e.g.,
][]{robins2010alternative, stensrud2022conditional,
  stensrud2022separable, robins2022interventionist}.\footnote{Some of
  this work uses the expression ``interventionist view of causality''
  to refer to their methods. The expression ``agency view of
  causality'' would have been more appropriate as all counterfactual
  causal inference involves interventions.} Briefly, these works
achieve identifiability of mediation parameters under an additional structural
assumption stating that treatment/exposure can be separated into
disjoint components whose effect operate independently through each of
the different causal pathways. Agency interventions are then
conceptualized on each separable component. Our goal in this section
is to define mediation parameters for the common case where such
structural information is unavailable, at the expense of abandoning
the agency interpretation of the path-specific effects.

  Consider the ATE $\psi = \E[Y(1) - Y(0)]$ introduced in
  \S\ref{sec:ate}, and note that under our assumed SCM we have
  $Y(a) = Y(a, Z(a), M(a, Z(a)))$. Then we could use the following
  counterfactuals to define path-specific effects:
\begin{align}
  Y_{S_0}&=Y(1, Z(1), M(1, Z(1))),\notag\\
  Y_{S_1}&=Y(0, Z(1), M(1, Z(1))),\notag\\
  Y_{S_2}&=Y(0, Z(0), M(1, Z(1))),\label{eq:decompnat}\\
  Y_{S_3}&=Y(0, Z(0), M(1, Z(0))),\notag\\
  Y_{S_4}&=Y(0, Z(0), M(0, Z(0))),\notag
\end{align}
where the causal effect operating through path $P_j$ is defined as
$\E[Y_{S_j-1} - Y_{S_j}]$. As before, the probability distribution of
$Y_{S_2}$ is not identified due to the recanting witness $Z$. However,
we can achieve an effect decomposition into path-specific effects
using information transfer interventions as follows. Let $Z_a$ denote
a random draw from the distribution of $Z(a)$ conditional on
$W$. Define

\begin{align*}
  Y_{S_1}' &= Y(0, Z(1), M(1, Z_1)),\\
  Y_{S_2}' &= Y(0, Z(0), M(1, Z_1)),\\
  Y_{S_2}'' &= Y(0, Z_0, M(1, Z(1))),\\
  Y_{S_3}'' &= Y(0, Z_0, M(1, Z(0))),
\end{align*}
where, in comparison to the definitions in (\ref{eq:decompnat}), we
have emulate the information transferred through some paths by means
of the random draws $Z_1$ and $Z_0$. For example, in $Y_{S_2}'$, the
effect of the action $A=1$ operating through the path
$A=1\to Z(1)\to M\to Y$ is emulated by the effect operating through a
path $A=1\to Z_1\to M\to Y$. This allows us to achieve the following
identifiable effect decomposition:

\begin{theorem}[Decomposition of the average treatment effect into
  path-specific effects]\label{theo:decompeffect}
  Define the path-specific causal effects as
  \begin{align*}
    \psi_{P_1} &= \E(Y_{S_0} - Y_{S_1})\\
    \psi_{P_2} &= \E(Y_{S_1}' - Y_{S_2}')\\
    \psi_{P_3} &= \E(Y_{S_2}'' - Y_{S_3}'')\\
    \psi_{P_4} &= \E(Y_{S_3}  - Y_{S_4}),
  \end{align*}
  as well as the parameter
  \[  \psi_{P_2\vee P_3} = E(Y_{S_1} - Y_{S_1}' + Y_{S_2}' -
    Y_{S_2}'' + Y_{S_3}'' - Y_{S_3}).\]
  Then we have the following decomposition of the
  average treatment effect $\psi=\E(Y_{S_0} - Y_{S_4})$:
  \[\psi = \psi_{P_1} + \psi_{P_2} +  \psi_{P_3} +
    \psi_{P_4} +  \psi_{P_2\vee P_3}.\]
\end{theorem}

As in the previous section, we have the following result showing that
this decomposition of the total causal effect satisfies the
path-specific null criterion:

\begin{theorem}\label{theo:sharpeffect}
  The contrasts defined in Theorem~\ref{theo:decompeffect} satisfy
  the path-specific sharp null criterion \ref{prop:sharp_ps} with
  respect to each path $P_j$.
\end{theorem}

Furthermore, in this case $\psi_{P_2\vee P_3}$ is also an effect
operating through either $A\to Z\ \to Y$ or $A\to Z\ \to M\to Y$,
which is equal to zero whenever there is no effect through at least
one of these paths (e.g., if $Z$ is not an intermediate confounder):

\begin{proposition}\label{prop:zeroeff}
  Let $\mathcal U$ denote the range of $U=(U_W, U_A, U_Z, U_M,
  U_Y)$. Assume $\mathcal U$ can be partitioned in sets
  $\mathcal U_1$, $\mathcal U_2$, and $\mathcal U_3$ such that the
  following hold almost surely
  \begin{itemize}
  \item $Z(1) = Z(0)$ in $\mathcal U_1$, and
  \item $\sup_{\bar z}|M(z_1) - Z(z_0)|=0$ in $\mathcal U_2$, and
  \item $\sup_{\bar z}|Y(z_1) - Y(z_0)|=0$ in $\mathcal U_3$.
  \end{itemize}Then $\psi_{P_2\vee P_3}=0$.
\end{proposition}

In order to present identification results for the above
decomposition, we will require the following overlap assumption which
guarantees that the functionals defined in
Theorem~\ref{theo:identifyeffect} are well defined.

\begin{assumptioniden}[Overlap assumption for decomposition of the ATE]\label{ass:overeffect}
  Assume the following hold for all $w$ such that $p(w)>0$
  \begin{itemize}
  \item $\p(a\mid w)>0$ for $a\in\{0,1\}$.
  \item $\p(z\mid A=1, w)>0$ implies $\p(z\mid A=0, w)>0$.
  \item $\p(m\mid A=1, z', w)>0$ implies $\p(m\mid A=0, z, w)>0$ for
    all $(z,z')$ such that $\p(z\mid a', w)>0$ and $\p(z'\mid a^\star,
    w)>0$ and for $a',a^\star\in \{0,1\}$.
  \end{itemize}
\end{assumptioniden}

\begin{theorem}[Identification of the path-specific decomposition of
  the average treatment effect]\label{theo:identifyeffect}
  Assume
  \ref{ass:overeffect}. Then, for $a'=1$ and $a^\star=0$, we have the
  following under the assumed SCM:
  \begin{align*}
    \E(Y_{S_0}) &= \E\{\E(Y\mid a', W)\},\\
    \E(Y_{S_1}) &= \E\left[\int\E(Y\mid a^\star, z,m,W)\dd\P(z, m\mid a',W)\right],\\
    \E(Y_{S_1}') &= \E\left[\int\E(Y\mid a^\star, z,m,W)\dd\P(z\mid a',
                   W)\dd\P(m\mid a',W)\right],\\
    \E(Y_{S_2}') = \E(Y_{S_2}'') &= \E\left[\int\E(Y\mid a^\star, z,m,W)\dd\P(z\mid a^\star,
                                   W)\dd\P(m\mid a',W)\right],\\
    \E(Y_{S_3}'') &= \E\left[\int\E(Y\mid a^\star, z,m,W)\dd\P(z\mid a^\star,
                    W)\dd\P(m\mid a',z',W)\dd\P(z'\mid
                    a^\star, W)\right],\\
    \E(Y_{S_3}) &= \E\left[\int\E(Y\mid a^\star, z,m,W)\dd\P(z\mid a^\star,
                  W)\dd\P(m\mid a',z,W)\right],\\
    \E(Y_{S_4}) &= \E\{\E(Y\mid a^\star, W)\}.
  \end{align*}
\end{theorem}
The proofs of the results in this section follow steps identical to
the proofs of the results in \S\ref{sec:pathanal}, where we note that
for binary $A$ the causal influence parameter
$\theta = \cov(A, Y-Y_S)$ and the ATE $\psi = \E[Y(1) - Y(0)]$ are
connected through the relation
$\theta = \E\{p(W)^2(1-p(W))^2\E[Y(1) - Y(0)\mid W]\}$, for
$p(w)=\P(A=1\mid W=w)$.

\section{Efficient estimation of the path-specific influence using the
  covariance contrast}\label{sec:estima}

In this section we discuss estimation of the covariance parameters
presented in Example~\ref{ex:covar}, and specifically the causal
influence decomposition in (\ref{eq:covdec}). Note that, to estimate
these covariances, it will be sufficient to construct an estimator of
the parameter $\tau^k_j = \E[f(A)Y_{S_j}^{(k)}]$ for a general
function $f$. Applications to $f(A)=A$ and $f(A)=1$ will allow us to
obtain estimates of $\E[AY_{S_j}^{(k)}]$ and $\E[Y_{S_j}^{(k)}]$,
which together with $\E[A]$ constitute the basis for computing the
covariance parameters $\E[AY_{S_j}^{(k)}] - \E[A]\E[Y_{S_j}^{(k)}]$.

Note also that estimation of $\E[f(A)Y_{S_0}^{(0)}] = \E[f(A)Y]$ and
$\E[A]$ is straightforward by means of empirical averages. Thus, in
the remaining of this section we focus on estimation of $\tau_j^k$
for
$(k,j)\in \{(0,1), (1,1),\allowbreak (1,2), (2,3), (0,3),
(0,4)\}$. For the sake of simplicity, we pursue estimators assuming
that $A$, $Z$, and $M$ are discrete random variables, although some of
the efficiency theory results of this section such as
Theorem~\ref{theo:vonmises} and the efficiency
bound hold for arbitrary variables.

The identification formulas in Theorem~\ref{theo:iden} provide a
relatively easy approach to obtain estimators of $\tau^k_j$. For
example, given any regression estimate of $\E[Y\mid A=a,M=m,Z=z,W=w]$
and an estimate of the probabilities $\P(M=m,Z=z,A=a\mid W=w)$, an
estimator may be constructed by plugging in these estimates in the
identification formulas. If the models for these regression and
probabilities are parametric and correctly specified, a maximum
likelihood plug-in estimation strategy is optimal in the sense that
the estimators converge to the optimal normal distribution at
$\sqrt{n}$-rate. However, if the model is non-parametric, a plug-in
estimation strategy results in first-order bias, which must be
corrected.

The general methods to characterize this first-order bias are rooted
in semi-parametric estimation theory
\citep[e.g.,][]{mises1947asymptotic, begun1983information, Bickel97,
  vanderVaart98, robins2009quadratic}, and in the theory for doubly
robust estimation using estimating equations
\citep{Robins00,Robins&Rotnitzky&Zhao94,vanderLaan2003,Bang05}. Under
this theory, the first-order bias is characterized in terms of the
so-called \textit{canonical gradient}. This canonical gradient also
characterizes the efficiency bound in the non-parametric model, and is
therefore also known as the \textit{efficient influence
  function}. Importantly, knowledge of the canonical gradient allows
the development of estimators under slow convergence rates for the
nuisance parameters involved. This is important because flexible
regression and estimation methods involving model selection must be
used when the model is non-parametric, and those flexible estimation
methods often fail to be consistent at parametric rate, though they
may be consistent at slower rates.

The following theorem illustrates the sense in which general plug-in
estimators are biased in the non-parametric model, and provides a
characterization of the bias in terms of the canonical gradient
$\varphi_j^k$. The specific formulas of the gradient for each
duple $(k,j)$ are useful to construct the estimators, but they are
cumbersome and somewhat uninformative, so we relegate their
presentation to the supplementary materials.

\begin{theorem}[First order von-Mises expansion]\label{theo:vonmises}
  Let $\tau_j^k(\G)$ denote the $\tau_j^k$ parameter evaluated
  at a distribution $\G$. Then, for any pair of distributions $\P$ and
  $\G$, we have the following:
  \[\tau_j^k(\G) - \tau_j^k(\P) =
    -\E_{\P}[\varphi_j^k(X;\G)] + R_j^k(\G,\P),\]
  where $R_j^k(\G,\P)$ is a second-order term of the form
  \[R_j^k(\G,\P)=\sum_{l}\int \omega(\G,\P)\{\kappa_l(\G) -
    \kappa_l(\P)\}\{\nu_l(\G) - \nu_l(\P)\}\dd\P,\] for functionals
  $\omega$, $\kappa$, and $\nu$ that vary with $(i,k)$. The specific
  form of each canonical gradient $\varphi_j^k$ and second order
  term $R_j^k(\G,\P)$ is given in the supplement.
\end{theorem}

An application of the above theorem with $\G$ equal to an estimate
$\hat\P$ of the true probability distribution $\P$ reveals that a
plug-in estimation strategy is generally biased in first-order, and
provides a representation of the bias as
$-\E_{\P}[\varphi_j^k(X;\hat\P)]$. Importantly, this theorem also
provides an avenue to correct for this first order bias. Specifically,
the first order bias may be estimated through the empirical average of
$\varphi_j^k(X_j;\hat\P)$ across observations $i$. If this bias is
added back to the plug-in estimator $\tau_j^k(\hat\P)$, one would
expect that the resulting estimator is unbiased in first-order. This
idea is formalized below in Theorem~\ref{theo:assnorm}.

The canonical gradient and the above reasoning are at the center of
several recent estimation methods that leverage machine learning and
flexible regression, such as the targeted learning framework
\citep{vdl2006targeted, vanderLaanRose11, vanderLaanRose18}, and
double machine learning \citep{chernozhukov2018double}. An important
feature of these approaches is the use of cross-fitting, which will
allow us to obtain $n^{1/2}$-convergence of our estimators while
avoiding entropy conditions that may be violated by data adaptive
estimators of the nuisance parameters \citep{zheng2011cross,
  chernozhukov2018double}. Let ${\cal P}_1, \ldots, {\cal P}_V$ denote
a random partition of the data set into $V$ prediction sets of
approximately the same size. That is,
${\cal P}_v\subset \{1, \ldots, n\}$;
$\bigcup_{j=1}^J {\cal P}_v = {\cal D}$; and
${\cal P}_v\cap {\cal P}_{v'} = \emptyset$. In addition, for each $v$,
the associated training sample is given by
${\cal T}_v = {\cal D} \setminus {\cal P}_v$. 


Our proposed estimator only requires estimation of the conditional
expectation of $Y$ given $(M, Z, A, W)$, which will be denoted by
$\m$, the probability mass function of $M$ given $(Z,A,W)$, denoted by
$\p_M$, the probability mass function of $Z$ given $(A,W)$, denoted by
$\p_Z$, and the probability mass function of $A$ given $W$, denoted by
$\p_A$. Let $\eta = (\m, \p_M, \p_Z,\p_Z)$, and note that the
canonical gradients $\varphi_j^k$ can be written as functions
$\varphi_j^k(X;\eta)$. Let $v(i)$ denote the prediction set to which
observation $i$ belongs, and let $\hat\eta_{v(i)}$ denote an estimator
of $\eta$ obtained using training data ${\cal T}_{v(i)}$. Then, the
estimator of $\tau_j^k$ is defined as
\[\hat\tau_j^k = \frac{1}{n}\sum_{i=1}^n\bar\varphi(X_j,\hat\eta_{v(i)}),\]
where $\bar\varphi$ is the uncentered canonical gradient given in the
supplementary materials.

The following theorem provides the conditions under which the above
estimator is expected to be efficient and asymptotically normal:
\begin{theorem}[Asymptotic linearity of the proposed
  estimator]\label{theo:assnorm}
  Assume that $R_j^k(\hat\eta,\eta) = o_\P(n^{-1/2})$ and that
  \begin{itemize}
  \item $\P\{\p_M(M=m\mid A,Z,W) < c_M\}=\P\{\hat\p_M(M=m\mid A,Z,W) <
    c_M\}=1$ for a constant $c_M$ and for all $m$,
  \item $\P\{\p_Z(Z=z\mid A,W) < c_Z\}=\P\{\hat\p_Z(Z=z\mid A,W) <
    c_Z\}=1$ for a constant $c_Z$, and that
  \item $\P\{\p_A(A=a\mid W) < c_A\}=\P\{\hat\p_A(A=a\mid W) <
    c_A\}=1$ for a constant $c_A$ and for all $a$.
  \end{itemize}
  Then
  \[\hat\tau_j^k - \tau_j^k
    =\frac{1}{n}\sum_{i=1}^n\varphi_j^k(X_i;\eta) + o_\P(n^{-1/2}).\]  
\end{theorem}
The proof of this theorem is sketched in the supplementary
materials. The arguments are standard in the analysis of estimators in
the targeted learning and double machine learning frameworks. The
above theorem implies that an estimator $\hat\theta_{P_j}$ of
$\theta_{P_j}$ constructed through an application of the above
estimator to $f(A)=A$ and $f(A)=1$ is also asymptotically linear.  An
application of the Delta method and the central limit theorem then
yields
\[n^{1/2}(\hat\theta_{P_j}-\theta_{P_j}) \rightsquigarrow
  N(0,\sigma^2),\] where $\sigma^2$ is the non-parametric efficiency
bound, which allows the construction of Wald-type confidence
intervals.



\section{Illustrations using real data}\label{sec:aplica}

In this section we apply the covariance path-specific decomposition
analysis proposed in the previous section to estimating path-specific
causal relations in two examples using publicly available data. In the
first example, we re-analyze the data from a recent study examining
gender differences in wage expectations among students at two Swiss
institutions of higher education \citep{fernandes2021gender}. In the
second example, we re-analyze data from a nationally representative
randomized experiment on how the framing of media discourse shapes
public opinion and immigration policy \citep{brader2008triggers}. The
datasets are publicly available in the \texttt{causalweight}
\citep{causalweight} and \texttt{mediation} \citep{mediation} R
packages, respectively. Code to reproduce our analyses is available at
\url{\github}. All nuisance parameters were estimated with extreme
gradient tree boosting where the hyperparameters are chosen from a
random grid of size 100 using the \textit{caret} \citep{caret} library in R. 

\subsection{Gender differences in wage expectations}
In this study, the authors administered a survey to 804 students at
the University of Fribourg and the University of Applied Sciences in
Bern in the year 2017. The survey contained a number of questions
regarding wage expectations, as well as a number of variables related
to wages such as the study program (business, economics,
communication, and business informatics), job or educational plans
after finishing the studies, the intended industry (trade, transport,
hospitality, communication, finance, etc.), as well as other variables
such as the age, parents education, nationality, and home
ownership. We study the causal relation between gender ($A$) and wage
expectations three years after graduation ($Y$) study program ($Z$)
and whether a student plans to continue obtaining further education or
work full time after graduation ($M$) as mediators. The outcome $Y$ is
recorded in a scale of 0-16, where 0 means less than 3500 Swiss Franc
(CHF) gross per month, 1 means 3500-4000 CHF, 2=4000-4500 CHF, etc.,
and 16=more than 11000 CHF. For simplicity and illustration purposes
we treat $Y$ as a numerical variable.

The results of our analysis are presented in Table~\ref{tab:wages}. We
conclude that most of the influence of gender on wages expectation 3
years after graduation operates through pathways independent of degree
and plans after study, which is a result similar to the original
result of \cite{fernandes2021gender}. Unlike the results in
\cite{fernandes2021gender}, which decompose the average treatment
effect into direct and indirect effects, the interpretation of the
estimates in Table~\ref{tab:wages} do not require to interpret
hypothetical and infeasible interventions that would modify someone's
gender. Instead, we compute the causal covariance between gender and
wage expectations $\theta$, and decompose it into path-specific
covariances.

\begin{table}[!htb]
  \centering
  \begin{tabular}{cccc}
    \hline
    Parameter & Estimate & Lower CI & Upper CI \\ 
    \hline
    $\theta$             & 0.330 & 0.240  & 0.421 \\ 
    $\theta_{P_1}$       & 0.302 & 0.205  & 0.398 \\ 
    $\theta_{P_2}$       & 0.021 & -0.285 & 0.327 \\ 
    $\theta_{P_3}$       & 0.003 & -0.004 & 0.010 \\ 
    $\theta_{P_4}$       & 0.003 & -0.003 & 0.010 \\ 
    $\theta_{P_2\vee P_3}$ & 0.001 & -0.332 & 0.335 \\ 
    \hline
  \end{tabular}
  \caption{Results in the gender and wage expectation illustrative
    example along with 95\% confidence intervals.}\label{tab:wages}
\end{table}

\subsection{Media discourse and public opinion and immigration policy}

In this study, the authors examine whether and how elite discourse
affects public opinion and action on immigration policy. They
conducted a randomized experiment in which 265 subjects are exposed to
different media stories about immigration. They employed a $2\times 2$
design in which they manipulate ethnic cues by altering the picture
and name of an immigrant featured in a hypothetical New York Times
story (white European vs Latin American). They also manipulate the
tone of the story, focusing on whether positive or negative
consequences of immigration, as well as conveying positive or negative
attitudes of governors and other citizens towards immigration. Our
treatment variable $A$ will take values 0, 1, or 2, with 0 denoting a
positive story about a white European, 1 denoting a negative story
about a white European immigrant or a positive story about a Latin
American immigrant, and 2 denoting a negative story about a Latin
American immigrant. The authors also collected information on age,
education, gender, and income of the study participants, which we
denote with $W$. A major hypothesis of the study was that anxiety and
is an important mediator of the causal influence of the framing of the
story on negative attitude towards immigration. Anxiety $M$ was
measured using a numerical scale from 3 to 12 where 3 indicates the
most negative feeling. Perceived harm $Z$ caused by immigration was
also measured in a scale between 2 and 8. The outcome of interest $Y$
is a four-point scale measuring a subject's attitude towards increased
immigration where larger values indicate more negative attitudes.

The results of the analysis are presented in
Table~\ref{tab:media}. These results largely agree with the results of
the original research article, with the difference that these analyses
provide more nuance in the sense that most of the influence of the
framing of the story on negative attitudes towards immigration is
mediated directly by anxiety through pathways that do not involve
perceived harm from immigration.

\begin{table}[!htb]
  \centering
  \begin{tabular}{cccc}
    \hline
    Parameter & Estimate & Lower CI & Upper CI \\ 
    \hline
    $\theta$             & 0.087  & 0.008  & 0.165  \\ 
    $\theta_{P_1}$       & 0.028  & -0.037 & 0.093  \\ 
    $\theta_{P_2}$       & 0.005  & -0.271 & 0.281  \\ 
    $\theta_{P_3}$       & 0.004  & -0.006 & 0.014  \\ 
    $\theta_{P_4}$       & 0.069  & 0.015  & 0.124  \\ 
    $\theta_{P_2\vee P_3}$& -0.020 & -0.290 & 0.251 \\ 
    \hline
  \end{tabular}
  \caption{Results in the media discourse illustrative example along with 95\% confidence intervals.}\label{tab:media}
\end{table}
\section{Discussion}\label{sec:discussion}

Our proposed approach to measuring causal influence by information
transfer interventions can be used in algorithmic fairness. As an
example, consider the methods proposed by \cite{nabi2019learning} to
learn fair optimal treatment policies. Their approach to fairness
relies on constructing a distribution where undesirable causal
pathways are ``removed'' and then learning optimal treatment policies
with respect to this fair but unobserved distribution. Their approach
to estimating a fair distribution relies on measuring the
path-specific effects, and then finding the closest (e.g., in
KL-divergence) distribution to the observed data distribution under
the constraint that these path-specific effects are null. Because our
proposed path-specific measures satisfy sharp null criteria, the
approach of \cite{nabi2019learning} to fairness could be adapted to
use our proposed path-specific effects to learn fair distributions.

Our proposal also shares connections to the general theory of causal
interventions presented by \cite{shpitser2016causal}. In their work,
the authors generalize and unify many interesting targets of causal
inference through the use of so-called edge- and path-interventions,
defined as interventions on the source node of the edge or path, where
the intervention operates only for the purpose of the specific
outgoing path. Our approach to achieving identification of
path-specific causal influence and path-specific causal effects also
entails a type of path-intervention. The difference with the approach
of \cite{shpitser2016causal} is that we do not restrict the
path-intervention to the source node, but instead intervene on nodes
in the other positions in the path. Specifically, by allowing
path-interventions to be defined in terms of the recanting witness
node, we are able to define path-specific effects which are
non-parametrically identifiable. A general theory for
path-intervention that allow for interventions on nodes other than the
source node seems to be an important direction of future work. The
interventions we propose also share connections with the interventions
proposed by \cite{malinsky2018intervening}. These general
``macro-level'' interventions focus on hypothetical worlds where the
structural equations had been replaced for alternative, user-given
equations. Some of our information transfer interventions can be
interpreted in this way, for example by noticing that the distribution
of the outcome in a hypothetical world where we set $A=\A$ is equal to
the distribution of the outcome where the argument $a$ is marginalized
out from the structural function $f_A$, where the marginalization is
with respect to the distribution $\P(a\mid W)$. 

Our discussion centers around a structural definition of causal
influence interpreted as the strength of dependence of the structural
functions on their arguments (see, e.g., Definition \ref{def:infl})
This criterion for causal influence may be too strong in some
cases. For example \cite{sprenger2018foundations} argues for a
probabilistic rather than structural definition of causal
influence. We conjecture that our proposed measures of causal
influence would also satisfy probabilistic null criteria, but leave
the proof of those results to future work.

\hlrev{Compared to the natural direct and indirect effects available
  in the literature, our proposal allows for the definition of
  parameters that are identified in the presence of a recanting
  witness. While changing the inferential target allowed us to bypass
  the recanting witness problem, this strategy comes at a
  price. Specifically, our parameters do not exactly decompose the
  total effect/influence into path-specific parameters. As detailed in
  Theorem~\ref{theo:decomp}, the decomposition we obtain involves an
  extra parameter that is equal to zero only in the absence of a
  recanting witness.}

Lastly, there are multiple interesting directions for future work that
build on the ideas presented in this paper. The first is that the
order of the decomposition we pursue where we proceed sequentially by
intervening on the paths $P_1$, $P_2$, $P_3$, and $P_4$ is
arbitrary. Other orderings for these paths can also be considered and
may be of more practical relevance in certain applications. The second
is that the ideas we present can be generalized to construct
path-specific effects for multiple ordered mediators and for
situations with time-varying treatments, mediators, and
covariates. The definition of path-specific parameters that measure
causal influence in those settings remains an open
problem. Furthermore, estimation of existing mediational parameters
such as interventional effects is notoriously hard in longitudinal
settings \cite[see e.g., ][]{diaz2022efficient} and we believe the
ideas introduced in this article can be useful to solve problems in
that setting.


\section*{Acknowledgments}
This work was supported through a Patient-Centered Outcomes Research
Institute (PCORI) Project Program Award (ME-2021C2-23636-IC).


\bibliographystyle{plainnat}
\bibliography{refs}
\end{document}


\maketitle
\section{Proofs of results in the paper}

\subsection{Proposition~\ref{theo:mono}}

\begin{proof}
  Assume without loss of generality that $\E(A)=0$. This follows from
  application of Lemma~\ref{lemma:final} below to $X=A$,
  $L = (W, U_M, U_Z, U_Y)$ and
  \[f(X, L) =\allowbreak f_Y(W, A, f_Z(W, A, U_Z), f_M(W, A, f_Z(W, A,
    U_Z), U_M), U_Y),\] after noticing that
  $\P(A\mid W, U_M, U_Z, U_Y) = \P(A\mid W)$ by assumption of the
  NPSEM, and that $\E[X(f(X, L) - \E(f(X,
  L)\mid L))=\cov(A, Y-\int Y(a)\dd\P(a\mid W))=\cov(A, Y- Y(\A))$.
\end{proof}

\begin{lemma}\label{lemma:pos}
  Let $(X,L)\sim \P$ any random variables, and let $f(x,y)$ be a
  function such that $f(\cdot,y)$ is increasing for all $y$. Assume
  that $\E(X)=0$. Define $(\rho,\beta) = \argmin R(g,b)$, where
  \[R(g,b)=\E[f(X, L) - g(L) - b X]^2.\]
  Then $\beta \geq 0$. Likewise, if $f(\cdot, y)$ is decreasing for all
  $y$, then $\beta \leq 0$.
\end{lemma}
\begin{proof}
  By standard arguments of projections in $L_2(P)$ spaces, it can be
  seen that $\rho(L) = \E[f(X, L) - b X\mid L]$ for any
  $b$. This yields
  \begin{align*}    
    R(\rho,\beta) &= \E\{[f(X,L)-\E(f(X,L)|L)]^2\}\\
    &-
    2\beta\E\{(X-\E(X\mid L))(f(X,L)-\E[f(X,L)\mid
      L])\}\\
                  &+\beta^2\E[\E^2(X\mid L)]\\
                  &= \E\{[f(X,L)-\E(f(X,L)|L)]^2\}\\
                  &-
                    2\beta\E\{(X-\E(X\mid L))(f(X,L)-f(\E(X\mid L),L))\}\\
                  &+\beta^2\E[\E^2(X\mid L)],
  \end{align*}
  where the second equality follows by adding and subtracting
  $f(\E(X\mid L), L)$ to the second factor of the second term. By
  monotonicity, the term $(X-\E(X\mid L))(f(X,L)-f(\E(X\mid L),L)$ is
  positive everywhere. Thus, if $\beta< 0$, then
  $R(\rho, |\beta|) < R(\rho, \beta)$, contradicting the definition of
  $\beta$ and proving the lemma.
\end{proof}

We also have the following general result

\begin{lemma}\label{lemma:final}
  Let $(X,L)\sim \P$ any random variables, and let $f(x,y)$ be a function such that
  $f(\cdot,y)$ is increasing for all $y$. Then $\E[X(f(X, L) - \E(f(X,
  L)\mid L))]\geq 0$.
\end{lemma}
\begin{proof}
  Continuing with the definition of $\beta$ in the above lemma, notice
  that $\E(X^2)\beta = \E[X(f(X, L) - \E(f(X,
  L)\mid L))]$, which proves the result.
\end{proof}
\subsection{Theorem~\ref{theo:total}}
\begin{proof}
We have
  \begin{align*}
    \P(Y_S\leq y\mid A=a) & = \int \P(Y_S\leq y\mid \A=a', W=w,
                            A=a)\dd\P_S(a', w\mid A=a)\\
                          & = \int \P(Y(a')\leq y\mid W=w,
                            A=a)\dd\P(a'\mid W=w)\dd\P(w\mid A=a)\\
                          & = \int \P(Y(a')\leq y\mid W=w,
                            A=a')\dd\P(a'\mid W=w)\dd\P(w\mid A=a)\\
                          & = \int \P(Y\leq y\mid W=w,
                            A=a')\dd\P(a'\mid W=w)\dd\P(w\mid A=a)\\
                          & = \int \P(Y\leq y\mid W=w)\dd\P(w\mid A=a),
  \end{align*}
  The first line follows by the law of iterated expectation, the
  second line by independence of $\A$ on all other data conditional on
  $W$ and by the fact that $\A$ is distributed as $A$ conditional on
  $W$, the third line follows by the assumption of the theorem, the
  fourth line because $Y(a')=Y$ in the event $A=a'$, and the last line
  by definition.
\end{proof}

\subsection{Theorem~\ref{theo:sharp}}

\begin{proof}
  The proof of Theorem~\ref{theo:sharp} proceeds as follows. We will
  prove the statement of the Theorem for each contrast of path
  separately.
  \begin{enumerate}
  \item For $P_1$, assume that $\P\{\sup_{\bar a}|Y(a_1, Z(a_2), M(a_3, Z(a_4))) -
    Y(a_0, Z(a_2), M(a_3, Z(a_4)))|=0\}=1$. Then we have
    \[ Y_{S_0}^{(0)} = Y(A, Z(A), M(A, Z(A)))
      = Y(\A, Z(A), M(A, Z(A)))  =Y_{S_1}^{(0)}\]
    almost surely, where the second equality follows by assumption. 
    
  \item For $P_2$, assume that
    $\P\{\sup_{\bar a, \bar m}|Y(a_1, Z(a_2), M(a_3, Z(a_4))) - Y(a_1,
    Z(a_0), M(a_3, Z(a_4)))|=0\}=1$. Then we have
    \begin{align*}
      \P(Y_{S_2}^{(1)}\leq  y\mid A=a) & = \P[Y_{S_2}^{(1)}\leq y\mid U\in
                                    \mathcal U_1, A=a]\P(U\in \mathcal U_1\mid
                                    A=a) \\
                                       & + \P[Y_{S_2}^{(1)}\leq y\mid U\in
                                    \mathcal U_2, A=a]\P(U\in \mathcal U_2\mid
                                    A=a),
    \end{align*}
    where $ \mathcal U_1$ and  $\mathcal U_2$ are the sets of the first
    statement of Lemma~\ref{lemma:aux} given below. We have \[Y_{S_2}^{(1)}=Y(\A,
      Z(\A), M(A, Z_A))=Y(\A, Z(A), M(A, Z_A))=Y_{S_1}^{(1)}\] a.s. in the event $U\in \mathcal U_1$. 
    We also have $Z(\A)=Z(A)$ a.s., and therefore $Y_{S_2}^{(1)}=Y_{S_1}^{(1)}$ in
    the event $U\in \mathcal U_2$.
  \item For $P_3$, assume that
    $\P\{\sup_{\bar a, \bar m}|Y(a_1, Z(a_2), M(a_3, Z(a_4))) - Y(a_1,
    Z(a_2), M(a_3, Z(a_0)))|=0\}=1$. Then we have
    \begin{align*}
      \P(Y_{S_3}^{(2)}\leq  y\mid A=a) & = \P[Y_{S_3}^{(2)}\leq y\mid U\in
                                     \mathcal U_1, A=a]\P(U\in \mathcal U_1\mid
                                     A=a) \\
                                       & + \P[Y_{S_3}^{(2)}\leq y\mid U\in
                                     \mathcal U_2, A=a]\P(U\in \mathcal U_2\mid
                                     A=a)\\
                                       & + \P[Y_{S_3}^{(2)}\leq y\mid U\in
                                     \mathcal U_3, A=a]\P(U\in \mathcal U_3\mid
                                     A=a),
    \end{align*}
    where $\mathcal U_1$, $\mathcal U_2$ and $\mathcal U_3$ are the
    sets of the second statement of Lemma~\ref{lemma:aux}. We have \[Y_{S_3}^{(2)}=Y(\A,
      Z_\A, M(A, Z(\A)))=Y(\A, Z_A, M(A, Z(A)))=Y_{S_2}^{(2)}\] a.s. in the event
    $U\in \mathcal U_1$. We also have $M(A, Z(\A))=M(A, Z(A))$ a.s. in the
    event  $U\in \mathcal U_2$, and therefore
    $Y_{S_3}^{(2)}=Y_{S_2}^{(2)}$. Lastly, in the event $U\in \mathcal U_3$ we
    have $Z(\A)=Z(A)$ a.s., and therefore $Y_{S_3}^{(2)}=Y_{S_2}^{(2)}$.
  \item For $P_4$, assume that
    $\P\{\sup_{\bar a, \bar m}|Y(a_1, Z(a_2), M(a_3, Z(a_4))) - Y(a_1,
    Z(a_2), M(a_0, Z(a_4)))|=0\}=1$. Then we have
    \begin{align*}
      \P(Y_{S_4}^{(0)}\leq  y\mid A=a) & = \P[Y_{S_4}\leq y\mid U\in
                                   \mathcal U_1, A=a]\P(U\in \mathcal U_1\mid
                                   A=a) \\
                                 & + \P[Y_{S_4}\leq y\mid U\in
                                   \mathcal U_2, A=a]\P(U\in \mathcal U_2\mid
                                   A=a),
    \end{align*}
    where $\mathcal U_1$ and $\mathcal U_2$ and $\mathcal U_3$ are the
    sets of the third statement of Lemma~\ref{lemma:aux}. We have \[Y_{S_4}^{(0)}=Y(\A,
      Z(\A), M(\A, Z(\A)))=Y(\A, Z(\A), M(A, Z(A)))=Y_{S_3}^{(0)}\] a.s. in the event
    $U\in \mathcal U_1$. We also have $M(\A, Z(\A))=M(A, Z(\A))$ a.s. in the
    event  $U\in \mathcal U_2$, and therefore
    $Y_{S_4}^{(0)}=Y_{S_3}^{(0)}$.
  \end{enumerate}
\end{proof}

\subsection{Theorem~\ref{theo:monopath}}
\begin{proof}
  The proof of this result follows steps analogous to those in the
  proof of Proposition~\ref{theo:mono}. Therefore, we will only
  sketch the arguments of the proof for $P_2$.  Assume $\E(A)=0$ and
  that $Y(a_1, Z(a_2), M(a_4, Z(a_3)))$ is increasing in $a_2$ for all
  $(a_1, a_3, a_4, u_Y,w)$. This implies that
  $Y(a_1, Z(a_2), M(a_4, z))$ is increasing in $a_2$ for all
  $(a_1, z, a_4, u_Y,w)$. Define $v_Y=(u_Y,u_M,u_Z)$, and let
  $(\rho, \beta_{P_2})=\argmin L_{P_2}(r, b)$, where
  \[ L_{P_2}(r, b) = \E\{Y(\bar A, Z(A), M(A, Z_A)) -
    r(W,M(A,Z_A),V_Y) - b A\}^2\] By Lemma~\ref{lemma:pos},
  $\beta_{P_2}\geq 0$. The proposition follows because, as before,
  $\E[\var(A\mid W)]\beta_{P_2} = \theta_{P_2} $, and therefore
  $\theta_{P_2}$ and $\beta_{P_2}$ have the same sign.
\end{proof}

\subsection{Theorem~\ref{theo:iden}}
\begin{proof}
  We will prove the results for each random variable at a time. First, note that

  \begin{align*}
    \P&(Y_{S_1}^{(0)}\leq y\mid A=a) = \\
      &=\int \P[Y(a',z,m)\leq y\mid A=a, Z(a)=z, 
        M(a,z)=m,W=w, \A=a']\times \\
      & \hspace{1cm}\dd\P(a'\mid w)\dd\P(Z(a)=z\mid a, 
        w)\dd\P(M(a,z)=m\mid Z(a)=z,a,w)\dd\P(w\mid a)\\
      &=\int \P[Y(a',z,m)\leq y\mid A=a, W=w]\dd\P(a'\mid w)\dd\P(z\mid a,
        w)\dd\P(m\mid z,a,w)\dd\P(w\mid a)\\
      &=\int \P[Y(a',z,m)\leq y\mid a',z,m, w]\dd\P(a'\mid w)\dd\P(z\mid a,
        w)\dd\P(m\mid z,a,w)\dd\P(w\mid a)\\
      &=\int \P[Y\leq y\mid a',z,m, w]\dd\P(a'\mid w)\dd\P(z\mid a,
        w)\dd\P(m\mid z,a,w)\dd\P(w\mid a),
  \end{align*}
  where the first equality follows by law of iterated expectation, the
  second one by the independence assumptions on the errors $U$ and
  definition of counterfactuals $M(a,z)$ and $Z(a)$, and the fourth
  one by definition of the counterfactual $Y(a,z,m)$.  Note that, to
  simplify notation, we removed the random variables from the right
  hand side of the $|$ symbol in the above probabilities.
  
  We also have
  \begin{align*}
    \P&(Y_{S_1}^{(1)}\leq y\mid A=a) = \\
      &=\int \P[Y(a',z,m)\leq y\mid A=a, Z(a)=z, 
        M(a,z')=m,W=w, Z_A = z', \A=a']\times \\
      & \hspace{1cm}\dd\P(a'\mid w)\dd\P(Z(a)=z\mid a,
        w)\dd\P(M(a,z')=m\mid Z(a)=z, a,w)\dd\P(z'\mid a, w)\dd\P(w\mid a)\\
      &=\int \P[Y(a',z,m)\leq y\mid A=a, W=w]\dd\P(a'\mid w)\dd\P(z\mid a,
        w)\dd\P(m\mid z',a,w)\dd\P(z'\mid a, w)\dd\P(w\mid a)\\
      &=\int \P[Y(a',z,m)\leq y\mid a',z,m, w]\dd\P(a'\mid w)\dd\P(z\mid a,
        w)\dd\P(m\mid z',a,w)\dd\P(z'\mid a, w)\dd\P(w\mid a)\\
      &=\int \P[Y\leq y\mid a',z,m, w]\dd\P(a'\mid w)\dd\P(z\mid a,
        w)\dd\P(m\mid a,w)\dd\P(w\mid a).
  \end{align*}

  Furthermore,
  \begin{align*}
    \P&(Y_{S_2}^{(1)}\leq y\mid A=a) = \\
      &=\int \P[Y(a',z,m)\leq y\mid A=a,Z(a')=z, 
        M(a,z')=m,W=w,Z_A=z', \A=a']\times \\
      & \hspace{1cm}\dd\P(a'\mid w)\dd\P(z'\mid a, w)\dd\P(M(a, z')=m\mid Z(a') = z,
        a, w)\dd\P(Z(a') = z\mid a, w)\dd\P(w\mid a)\\
      &=\int \P[Y(a',z,m)\leq y\mid A=a,W=w]\dd\P(a'\mid w)\dd\P(z'\mid a, w)\dd\P(m\mid z',
        a, w)\dd\P(z\mid a', w)\dd\P(w\mid a)\\
      &=\int \P[Y\leq y\mid a',z,m,w]\dd\P(a'\mid w)\dd\P(z'\mid a, w)\dd\P(m\mid z',
        a, w)\dd\P(z\mid a', w)\dd\P(w\mid a)\\
      &=\int \P[Y\leq y\mid a',z,m,w]\dd\P(a'\mid w)\dd\P(m\mid
        a, w)\dd\P(z\mid a', w)\dd\P(w\mid a)\\
      &=\int \P[Y\leq y\mid a',z,m,w]\dd\P(z,a'\mid w)\dd\P(m, w\mid
        a)\\
  \end{align*}

  and,
  \begin{align*}
    \P&(Y_{S_2}^{(2)}\leq y\mid A=a) = \\
      &=\int \P[Y(a',z,m)\leq y\mid A=a,Z(a)=z', 
        M(a,z')=m,W=w,Z_\A=z, \A=a']\times \\
      & \hspace{1cm}\dd\P(a'\mid w)\dd\P(z\mid a', w)\dd\P(M(a, z')=m\mid Z(a) = z',
        a, w)\dd\P(Z(a) = z'\mid a, w)\dd\P(w\mid a)\\
      &=\int \P[Y(a',z,m)\leq y\mid A=a,W=w]\dd\P(a'\mid w)\dd\P(z\mid a', w)\dd\P(m\mid z',
        a, w)\dd\P(z'\mid a, w)\dd\P(w\mid a)\\
      &=\int \P[Y\leq y\mid a',z,m,w]\dd\P(a'\mid w)\dd\P(z\mid a', w)\dd\P(m\mid z',
        a, w)\dd\P(z'\mid a, w)\dd\P(w\mid a)\\
      &=\int \P[Y\leq y\mid a',z,m,w]\dd\P(z,a'\mid w)\dd\P(m, w\mid
        a)\\
  \end{align*}
  
  In addition,
  \begin{align*}
    \P&(Y_{S_3}^{(2)}\leq y\mid A=a) = \\
      &=\int \P[Y(a',z,m)\leq y\mid A=a,Z(a')=z', 
        M(a,z')=m,W=w,Z_\A=z, \A=a']\times \\
      & \hspace{1cm}\dd\P(z\mid a', w)\dd\P(M(a, z')=m\mid Z(a') = z',
        a, w)\dd\P(Z(a') = z'\mid a, w)\dd\P(a'\mid w)\dd\P(w\mid a)\\
      &=\int \P[Y(a',z,m)\leq y\mid A=a,W=w]\dd\P(z\mid a', w)\dd\P(m\mid z',
        a, w)\dd\P(z'\mid a', w)\dd\P(a'\mid w)\dd\P(w\mid a)\\
      &=\int \P[Y\leq y\mid a',z,m,w]\dd\P(z\mid a', w)\dd\P(m\mid z',
        a, w)\dd\P(z'\mid a', w)\dd\P(a'\mid w)\dd\P(w\mid a),
  \end{align*}

  As well as,
  \begin{align*}
    \P&(Y_{S_3}^{(0)}\leq y\mid A=a) = \\
      &=\int \P[Y(a',z,m)\leq y\mid A=a,Z(a')=z, 
        M(a,z)=m,W=w, \A=a']\times \\
      & \hspace{1cm}\dd\P(a'\mid w)\dd\P(M(a, z)=m\mid Z(a') = z,
        a, w)\dd\P(Z(a') = z\mid a, w)\dd\P(w\mid a)\\
      &=\int \P[Y(a',z,m)\leq y\mid A=a,W=w]\dd\P(a'\mid w)\dd\P(m\mid z,
        a, w)\dd\P(z\mid a', w)\dd\P(w\mid a)\\
      &=\int \P[Y\leq y\mid a',z,m,w]\dd\P(a'\mid w)\dd\P(m\mid z,
        a, w)\dd\P(z\mid a', w)\dd\P(w\mid a)\\
      &=\int \P[Y\leq y\mid a',z,m,w]\dd\P(z,a'\mid w)\dd\P(m\mid z,
        a, w)\dd\P(w\mid a),
  \end{align*}

  Lastly,
  \begin{align*}
    \P&(Y_{S_4}^{(0)}\leq y\mid A=a) = \\
      &=\int \P[Y(a',z,m)\leq y\mid A=a, Z(a')=z,
        M(a',z)=m,W=w, \A=a']\times \\
      & \hspace{1cm}\dd\P(a'\mid w)\dd\P(M(a',z)=m\mid
        Z(a')=z,a,w)\dd\P(Z(a')=z\mid a, w)\dd\P(w\mid a)\\
      &=\int \P[Y(a',z,m)\leq y\mid A=a, W=w]\dd\P(a'\mid
        w)\dd\P(m\mid z,a',w)\dd\P(z\mid a',w)\dd\P(w\mid a)\\
      &=\int \P[Y(a',z,m)\leq y\mid m,a,z, w]\dd\P(a'\mid w)\dd\P(m\mid z,a',w)\dd\P(z\mid a',w)\dd\P(w\mid a)\\
      &=\int \P[Y\leq y\mid m,a',z, w]\dd\P(a'\mid w)\dd\P(m\mid
        z,a',w)\dd\P(z\mid a',w)\dd\P(w\mid a)\\
      &=\int \P[Y\leq y\mid w]\dd\P(w\mid a).
  \end{align*}

\end{proof}

\subsection{Proposition~\ref{prop:zero}}
First, note that in the event $U\in \mathcal U_1$ we have
$Y_{S_1}^{(0)}=Y_{S_3}^{(0)}$, $Y_{S_2}^{(2)}=Y_{S_3}^{(2)}$, and $Y_{S_1}^{(1)}=Y_{S_2}^{(1)}$. In
the event $U\in \mathcal U_2$ we have $Y_{S_1}^{(0)}=Y_{S_1}^{(1)}$,
$Y_{S_2}^{(2)}=Y_{S_3}^{(2)}$, and $Y_{S_2}^{(1)}=Y_{S_3}$. Furthermore, in the
event $U\in \mathcal U_3$ we have $Y_{S_1}^{(0)}=Y_{S_2}^{(2)}$,
$Y_{S_1}^{(1)}=Y_{S_2}^{(1)}$, and $Y_{S_3}^{(2)}=Y_{S_3}^{(0)}$.

We have  
\begin{align}
  D(\P_{S_1}^{(0)}, \P_{S_1}^{(1)})   &=D(\P_{S_3}^{(0)}(\cdot\mid
                           \mathcal U_1), \P_{S_2}^{(1)}(\cdot\mid
                           \mathcal U_1))\P(\mathcal U_1)\label{eq:1}\\
                         &+  D(\P_{S_2}^{(2)}(\cdot\mid
                           \mathcal U_3), \P_{S_2}^{(1)}(\cdot\mid
                           \mathcal U_3))\P(\mathcal U_3).\label{eq:4}\\
  D(\P_{S_2}^{(1)}, \P_{S_2}^{(2)})  &=D(\P_{S_2}^{(1)}(\cdot\mid
                           \mathcal U_1), \P_{S_3}^{(2)}(\cdot\mid
                              \mathcal U_1))\P(\mathcal U_1) \label{eq:2}\\
                                      &+ D(\P_{S_3}^{(0)}(\cdot\mid
                           \mathcal U_2), \P_{S_3}^{(2)}(\cdot\mid
    \mathcal U_2))\P(\mathcal U_2) \label{eq:6}\\
  &+ D(\P_{S_2}^{(1)}(\cdot\mid
                           \mathcal U_3), \P_{S_2}^{(2)}(\cdot\mid
    \mathcal U_3))\P(\mathcal U_3).\label{eq:5}\\
  D(\P_{S_3}^{(2)}, \P_{S_3}^{(0)})  &=D(\P_{S_3}^{(2)}(\cdot\mid
                                       \mathcal U_1), \P_{S_3}^{(0)}(\cdot\mid
                             \mathcal U_1))\P(\mathcal U_1) \label{eq:3}\\
                           &+ D(\P_{S_3}^{(2)}(\cdot\mid
                             \mathcal U_2), \P_{S_3}^{(0)}(\cdot\mid
                             \mathcal U_2))\P(\mathcal U_2)\label{eq:7}.
\end{align}
By the assumptions of the lemma we have
$(\ref{eq:1})+(\ref{eq:2})+(\ref{eq:3})=0$, $(\ref{eq:4})+(\ref{eq:5})=0$,
and $(\ref{eq:6})+(\ref{eq:7})=0$, concluding the proof of the lemma.

\section{Theorem~\ref{theo:vonmises}}
We first give the form of the canonical gradients, and then prove the
result. The canonical gradients are given by
$\varphi_j^{(k)}(X;\P)=\bar\varphi_1^{(0)}(X;\P)-\tau_j^k(\P)$, where
\begin{align*}
  \bar\varphi_1^{(0)}(X;\P)
  & = \frac{h_1^{(0)}(M,Z,W)}{\p(M,Z\mid A,W)}\{Y-\E(Y\mid A,Z,M,W)\} \\
  & +\int f(a)\E(Y\mid A,z,m,W) \dd\P(m,z,a\mid W)                               \\
  & -\int f(a)\E(Y\mid a',z,m,W)\dd\P(m,z,a\mid W)\dd\P(a'\mid W)\\
  & +f(A)\int \E(Y\mid a', Z, M, W)\dd\P(a'\mid W)             \\
  \varphi_1^{(1)}(X;\P)
  & =\frac{h_1^{(1)}(M,Z,W)}{\p(M,Z\mid A,W)}\{Y-\E(Y\mid A,Z,M,W)\}  \\
  & +\int f(a)\E(Y\mid A,z,m,W)\dd\P(m\mid a, W)\dd\P(z\mid a,
    W)\dd\P(a\mid W)\\
  & -\int f(a)\E(Y\mid a',z,m,W)\dd\P(m\mid a, W) \dd\P(z\mid a,
    W)\dd\P(a\mid W)\dd\P(a'\mid W) \\
  & +f(A)\int\E(Y\mid a', Z, m, W)\dd\P(m\mid A, W)\dd\P(a'\mid  W)\\
  & -f(A)\int\E(Y\mid a', z, m, W)\dd\P(m\mid A, W)\dd\P(z\mid A, W) \dd\P(a'\mid W)\\  
  & +f(A)\int \E(Y\mid a',z,M, W)\dd\P(z\mid A,W)\dd\P(a'\mid W) \\
  \bar\varphi_2^{(1)}(X;\P)
  & =\frac{h_2^{(1)}(M,W)}{\p(M\mid Z,A, W)}\{Y-\E(Y\mid A,Z,M,W)\}  \\
  & +\int f(a)\E(Y\mid A,Z,m,W)
    \dd\P(m\mid a, W)\dd\P(a\mid W)                                                           \\
  & -\int f(a)\E(Y\mid a',z,m,W)\dd\P(m\mid a, W)\dd\P(a\mid W)\dd\P(z, a'\mid W) \\
  & +f(A)\int \E(Y\mid a',z,M, W)\dd\P(z, a'\mid W)\\
  \bar\varphi_3^{(2)}(X;\P)
  & =\frac{h_3^{(2)}(M,A,W)}{\p(M\mid Z,A W)}\{Y-\E(Y\mid
    A,Z,M,W)\}                                                                  \\
  & +\int f(a)\E(Y\mid A, Z, m, W)\dd\P(m\mid a, z', W)\dd\P(z'\mid A,
    W)\dd\P(a\mid W)                                                            \\
  & -\int f(a)\E(Y\mid a', z, m, W)\dd\P(m\mid a, z', W)\dd\P(z'\mid
    a', W)\dd\P(z\mid a', W)\dd\P(a\mid W)\dd\P(a'\mid W) \\
  &+ \frac{f(A)}{\p(Z\mid A, W)}\int \E(Y\mid a', z, M, W)\dd\P(z\mid a', W) \p(Z\mid a',
    W)\dd\P(a'\mid W)\\
  &- \frac{f(A)}{\p(Z\mid A, W)}\int \E(Y\mid a', z, m,
    W)\dd\P(m\mid A, Z, W)\dd\P(z\mid a', W) \p(Z\mid a',
    W)\dd\P(a'\mid W)\\
  & +\int f(a)\E(Y\mid A, z, m, W)\dd\P(z\mid A, W)\dd\P(m\mid a, Z,
    W)\dd\P(a\mid W) \\
  & -\int f(a)\E(Y\mid A, z, m, W)\dd\P(m\mid a, z',
    W)\dd\P(z\mid A, W)\dd\P(z'\mid A, W)\dd\P(a\mid W) \\
  & + f(A)\int \E(Y\mid a', z, m, W)\dd\P(m\mid A, z',
    W)\dd\P(z\mid a', W)\dd\P(z'\mid a', W)\dd\P(a'\mid W)\\
  \bar\varphi_3^{(0)}(X;\P)
  & =\frac{h_3^{(0)}(M,Z,W)}{\p(M\mid Z,A, W)}\{Y-\E(Y\mid
    A,Z,M,W)\}\\
  & + \frac{f(A)}{\p(Z\mid A, W)}\int \E(Y\mid a', Z, M, W)\p(Z\mid
    a', W)\dd\P(a'\mid W)\\
  & + \frac{f(A)}{\p(Z\mid A, W)}\int \E(Y\mid a', Z, m, W)\dd\P(m\mid
    A, Z, W)\p(Z\mid
    a', W)\dd\P(a'\mid W)\\
  & +\int f(a)\E(Y\mid A, Z, m, W)\dd\P(m\mid a, Z, W)\dd\P(a\mid W)\\
  & -\int f(a)\E(Y\mid a', z, m, W)\dd\P(m\mid a, z, W)\dd\P(z\mid a',
    W)\dd\P(a\mid W)\dd\P(a'\mid W)\\
  &+ f(A)\int \E(Y\mid a', z, m, W)\dd\P(m\mid A, z, W)\dd\P(z\mid a',
    W)\dd\P(a'\mid W)\\
  \bar\varphi_4^{(0)}(X;\P)
  & = h_4^{(0)}(W)\{Y - \E(Y\mid W)\} + f(A)\E(Y\mid W)
\end{align*}
where
\begin{align*}
  h_1^{(0)}(M,Z,W)&=\int f(a) \p(M,Z\mid a, W)\dd\P(a\mid W)\\
  h_1^{(1)}(M,Z,W)&=\int f(a)\p(Z\mid a, W)\p(M\mid a, W)\dd\P(a\mid W)\\
  h_2^{(1)}(M,W)&=\int f(a)\p(M\mid a, W)\dd\P(a\mid W)\\
  h_3^{(2)}(M,A,W)&=\int f(a)\p(M\mid a, z',W)\dd\P(z'\mid A,
                    W)\dd\P(a\mid W)\\
  h_3^{(0)}(M,Z,W)&=\int f(a)\p(M\mid a, Z,W)\dd\P(a\mid W)\\
  h_4^{(0)}(W)&=\int f(a)\dd\P(a\mid W).
\end{align*}

\begin{proof}
  The proof of the theorem proceeds as follows. We prove the result
  for $(i,k)=(0,1)$ and $(i,k)=(1,1)$, the other results can be
  obtained using the same arguments. We assume the distributions $\G$
  and $\P$ have densities $\g$ and $\p$ dominated by a measure
  $\lambda$. We use $\int f\dd\lambda$ to denote
  $\int f(x)\dd\lambda(x)$, and use
  $\int bf'g^\star h'^\star\dd\lambda$ to denote
  $\int b(m,z,a,w)f(m,z,a',w)\allowbreak g(m, z^\star,a,w)h(m,z^\star,
  a', w)\lambda(m,z,z^\star, a, a', w)$.

  For $(i,k)=(0,1)$, we have
  \begin{align*}
    \tau_1^{(0)}&(\G) - \tau_1^{(0)}(\P) +
                  \E_\P[\varphi_1^{(0)}(X;\G)]=\\
    &=\int f\frac{\g_M\g_Z\g_A}{\g_M'\g_Z'}(\m_\P' -
      \m_\G')\p_M'\p_Z'\p_A'\p_W\dd\lambda\\
                &+\int f\m_\G'\g_M\g_Z\g_A\p_A'\p_W\dd\lambda - \int
                  f\m_\G'\g_M\g_Z\g_A\g_A'\p_W\dd\lambda\\
                &+\int f\m_\G'\p_M\p_Z\p_A\g_A'\p_W\dd\lambda - \int
                  f\m_\G'\p_M\p_Z\p_A\p_A'\p_W\dd\lambda\\
    &=\int
      f\left(\frac{\p_M'\p_Z'}{\g_M'\g_Z'}-1\right)\left(\m_\P'-\m_\G'\right)\g_M\g_Z\g_A\p_A'\p_W\dd\lambda\\
    &+\int f\left(\g_M\g_Z\g_A - \p_M\p_Z\p_A\right)\left(\m_\P\p_A'-\m_\G\g_A'\right)\p_W\dd\lambda.
  \end{align*}
  Likewise, for $(i,k)=(1,1)$, we have
  \begin{align*}
    \tau_1^{(1)}&(\G) - \tau_1^{(1)}(\P) +
                  \E_\P[\varphi_1^{(1)}(X;\G)]=\\
                &=\int f\frac{\g_M^\star\g_Z^\star\g_Z\g_A}{\g_M'\g_Z'}(\m_\P' -
                  \m_\G')\p_M'\p_Z'\p_A'\p_W\dd\lambda\\
                &+\int f\m_\G'\g_M^\star\g_Z^\star\g_Z\g_A\p_A'\p_W\dd\lambda - \int
                  f\m_\G'\g_M^\star\g_Z^\star\g_Z\g_A\g_A'\p_W\dd\lambda\\
                &+\int f\m_\G'\g_M^\star\g_Z^\star\p_Z\g_A\g_A'\p_W\dd\lambda - \int
                  f\m_\G'\g_M^\star\g_Z^\star\g_Z\g_A\g_A'\p_W\dd\lambda\\
                &+\int f\m_\G'\p_M^\star\p_Z^\star\g_Z\p_A\g_A'\p_W\dd\lambda - \int
                  f\m_\P'\p_M^\star\p_Z^\star\p_Z\p_A\p_A'\p_W\dd\lambda\\
                &=\int f\left(\frac{\p_M'\p_Z'}{\g_M'\g_Z'}-1\right)(\m_\P' -
                  \m_\G')\g_M^\star\g_Z^\star\g_Z\g_A\p_A'\p_W\dd\lambda\\
                &+\int f\m_\P'\g_M^\star\g_Z^\star\g_Z\g_A\p_A'\p_W\dd\lambda - \int
                  f\m_\G'\g_M^\star\g_Z^\star\g_Z\g_A\g_A'\p_W\dd\lambda\\
                &+\int f\m_\G'\g_M^\star\g_Z^\star\p_Z\g_A\g_A'\p_W\dd\lambda - \int
                  f\m_\G'\g_M^\star\g_Z^\star\g_Z\g_A\g_A'\p_W\dd\lambda\\
                &+\int f\m_\G'\p_M^\star\p_Z^\star\g_Z\p_A\g_A'\p_W\dd\lambda - \int
                  f\m_\P'\p_M^\star\p_Z^\star\p_Z\p_A\p_A'\p_W\dd\lambda \\
                &+\int
                  f\m_\G'\p_M^\star\p_Z^\star\p_Z\p_A\g_A'\p_W\dd\lambda -
                  \int
                  f\m_\G'\p_M^\star\p_Z^\star\p_Z\p_A\g_A'\p_W\dd\lambda\\
                &=\int f\left(\frac{\p_M'\p_Z'}{\g_M'\g_Z'}-1\right)(\m_\P' -
                  \m_\G')\g_M^\star\g_Z^\star\g_Z\g_A\p_A'\p_W\dd\lambda\\
                &+\int f(\m_\P'\p_A' - \m_\G'\g_A')\g_M^\star\g_Z^\star\g_Z\g_A\p_W\dd\lambda\\
                &+\int f\m_\G'\g_M^\star\g_Z^\star(\p_Z-\g_Z)\g_A\g_A'\p_W\dd\lambda\\
                &+\int f\m_\G'\p_M^\star\p_Z^\star(\g_Z -\p_Z)\p_A\g_A'\p_W\dd\lambda\\
                &+\int
                  f(\m_\G'\g_A'-\m_\P'\p_A')\p_M^\star\p_Z^\star\p_Z\p_A\p_W\dd\lambda\\
                &=\int f\left(\frac{\p_M'\p_Z'}{\g_M'\g_Z'}-1\right)(\m_\P' -
                  \m_\G')\g_M^\star\g_Z^\star\g_Z\g_A\p_A'\p_W\dd\lambda\\
                &+\int f(\m_\P'\p_A' - \m_\G'\g_A')(\g_M^\star\g_Z^\star\g_Z\g_A-\p_M^\star\p_Z^\star\p_Z\p_A)\p_W\dd\lambda\\
                &+\int f\m_\G'(\g_M^\star\g_Z^\star\g_A-\p_M^\star\p_Z^\star\p_A)(\p_Z-\g_Z)\g_A'\p_W\dd\lambda
  \end{align*}

\end{proof}

\subsection{Proof of Theorem \ref{theo:assnorm}}
\begin{proof}
  To simplify notation we remove the index $j$ and superindex $k$ from
  $\tau_j^k$ to simplify notation. Let $\Pnj$ denote the empirical
  distribution of the prediction set ${\cal P}_v$, and let $\Gnj$
  denote the associated empirical process $\sqrt{n/J}(\Pnj-\P)$. Note
  that
  \[\hat\tau = \frac{1}{V}\sum_{v=1}^V\Pnj \bar\varphi(\cdot;\hat
      \eta_v),\,\,\,\tau=\P \bar\varphi(\cdot;\eta).\]Thus,
  \[  \sqrt{n}\{\hat\tau - \tau\}=\Gn \{\bar\varphi(\cdot;\eta)
    - \tau\} + R_{n,1} + R_{n,2},\]
  where
  \[  R_{n,1}  =\frac{1}{\sqrt{V}}\sum_{v=1}^V\Gnj\{\bar\varphi(\cdot;\hat
      \eta_{v}) - \bar\varphi(\cdot;\eta)\},\,\,\,
    R_{n,2}  = \frac{\sqrt{n}}{J}\sum_{v=1}^V\P\{\bar\varphi(\cdot;\hat
      \eta_v)-\tau\}.
  \]
  It remains to show that $R_{n,1}$ and $R_{n,2}$ are $o_P(1)$.
  Theorem \ref{theo:vonmises} together with the assumption that
  $R_j^{(k)}(\hat\eta,\eta) = o_\P(n^{-1/2})$ shows that $R_{n,2}=o_\P(1)$. For
  $R_{n,1}$ we use empirical process theory to argue
  conditional on the training sample ${\cal T}_v$. In particular,
  Lemma 19.33 of \cite{vanderVaart98} applied to the class of
  functions ${\cal F} = \{\bar\varphi(\cdot;\hat \eta_v) - \bar\varphi(\cdot;\eta)\}$
  (which consists of one element) yields
  \[E\left\{\big|\Gnj (\bar\varphi(\cdot;\hat \eta_v) -
      \bar\varphi_j^k(\cdot;\eta))\big| \,\bigg|\, {\cal
        T}_v\right\}\lesssim \frac{2C\log 2}{n^{1/2}} +
    ||\bar\varphi(\cdot;\hat \eta_v) - \bar\varphi(\cdot;\eta)||(\log
    2)^{1/2}\] The assumption
  $R_j^{(k)}(\hat\eta,\eta) = o_\P(n^{-1/2})$ can only hold if
  $\hat\eta-\eta=o_\P(1)$, therefore the right hand side is
  $o_\P(1)$. Lemma 6.1 of \cite{chernozhukov2018double} may now be
  used to argue that conditional convergence implies unconditional
  convergence, concluding the proof.

\end{proof}

\section{Auxiliary results}
\begin{lemma}\label{lemma:aux}
  Let $\mathcal U$ denote the range of $U=(U_W, U_A, U_Z, U_M,
  U_Y)$. The following statements are true:
  \begin{enumerate}
  \item $\P\{\sup_{\bar a}|Y(a_1, Z(a_2), M(a_3, Z(a_4))) -
    Y(a_1, Z(a_0), M(a_3, Z(a_4)))|=0\}=1$ implies $\mathcal U$ can be
    partitioned into sets
    $\mathcal U_1$ and $\mathcal U_2$ such that
    \begin{enumerate}
    \item $\P\{\sup_{\bar a,
      \bar z}|Y(a_1, z_1, M(a_3, Z(a_4))) - Y(a_1, z_2, M(a_3, Z(a_4)))|=0\mid
      U\in \mathcal U_1\}=1$, and
    \item $\P\{\sup_{\bar a}|Z(a_2) - Z(a_0)|=0\mid
      U\in \mathcal U_2\}=1$,
    \end{enumerate}    
    where $z_1\in\supp\{Z(a_2)\}$ and $z_2\in\supp\{Z(a_0)\}$.
  \item $\P\{\sup_{\bar a}|Y(a_1, Z(a_2), M(a_3, Z(a_4))) -
    Y(a_1, Z(a_2), M(a_3, Z(a_0)))|=0\}=1$ implies $\mathcal U$ can be
    partitioned into sets
    $\mathcal U_1$, $\mathcal U_2$, and $\mathcal U_3$ such that
    \begin{enumerate}
    \item $\P\{\sup_{\bar a,
        \bar m}|Y(a_1, Z(a_2), m_1) - Y(a_1, Z(a_2), m_2)|=0\mid
      U\in \mathcal U_1\}=1$, and
    \item $\P\{\sup_{\bar a, \bar z}|M(a_3, z_1) - M(a_3, z_2)|=0\mid
      U\in \mathcal U_2\}=1$, and
    \item $\P\{\sup_{\bar a}|Z(a_4) - Z(a_0)|=0\mid
      U\in \mathcal U_3\}=1$
    \end{enumerate}    
    where $m_1\in\supp\{M(a_3, Z(a_4))\}$ and $m_2\in\supp\{M(a_3,
    Z(a_0))\}$, and  $z_1\in\supp\{Z(a_4)\}$ and
    $z_2\in\supp\{Z(a_0)\}$.
  \item $\P\{\sup_{\bar a}|Y(a_1, Z(a_2), M(a_3, Z(a_4))) -
    Y(a_1, Z(a_2), M(a_0, Z(a_4)))|=0\}=1$ implies $\mathcal U$ can be
    partitioned into sets
    $\mathcal U_1$ and $\mathcal U_2$ such that
    \begin{enumerate}
    \item $\P\{\sup_{\bar a,
        \bar m}|Y(a_1, Z(a_2), m_1) - Y(a_1, Z(a_2), m_2)|=0\mid
      U\in \mathcal U_1\}=1$, and\label{state:1}
    \item $\P\{\sup_{\bar a}|M(a_3, Z(a_4)) - M(a_0, Z(a_4))|=0\mid
      U\in \mathcal U_2\}=1$,\label{state:2}
    \end{enumerate}    
    where $m_1\in\supp\{M(a_3, Z(a_4))\}$, and
    $m_2\in\supp\{M(a_0, Z(a_4))\}$.
  \end{enumerate}
\end{lemma}
\begin{proof}
  We show the proof for the third statement, assuming a discrete
  variable $M$. The other statements can be proved using parallel
  arguments. The statement is of the form \textit{if p then q}, we
  will prove the contrapositive \textit{if not q then not p}. We have
    \begin{multline*}|Y(a_1, Z(a_2), M(a_3, Z(a_4))) -
      Y(a_1, Z(a_2), M(a_0, Z(a_4)))| =\\ \sum_{m\neq m'}|Y(a_1, Z(a_2), m) -
      Y(a_1, Z(a_2), m')|\one\{M(a_3, Z(a_4))=m, M(a_0, Z(a_4))=m'\}.
    \end{multline*}
    Assume there is a set $\mathcal U^\star\subseteq \mathcal U$ with
    $\P(\mathcal U^\star)>0$ and values $\bar a$,
    $m_1\in\supp\{M(a_3, Z(a_4))\}$, and
    $m_2\in\supp\{M(a_0, Z(a_4))\}$ such that
    $Y(a_1, Z(a_2), m_1) \neq Y(a_1, Z(a_2), m_2)$ and
    $M(a_3, Z(a_4))\neq M(a_0, Z(a_4))$ in $\mathcal U^\star$ (i.e.,
    assume \textit{not q}). Then,
    \begin{multline*}|Y(a_1, Z(a_2), M(a_3, Z(a_4))) -
      Y(a_1, Z(a_2), M(a_0, Z(a_4)))| \geq\\ |Y(a_1, Z(a_2), m_1) -
      Y(a_1, Z(a_2), m_2)|\one\{M(a_3, Z(a_4))=m_1, M(a_0,
      Z(a_4))=m_2\}.
    \end{multline*}
    The probability that the term in the right hand side is positive is
    greater than zero in $\mathcal U^\star$ (i.e., \textit{not p}),
    concluding the proof of the claim.    
  \end{proof}
  
\bibliographystyle{plainnat}
\bibliography{refs}